\documentclass[twocolumn]{aastex61}
\pdfoutput=1 
\usepackage{amsmath,amstext}
\usepackage[T1]{fontenc}
\usepackage{apjfonts} 
\usepackage[figure,figure*]{hypcap}
\usepackage{scrextend}
\usepackage{footnote}
\usepackage{hyperref}
\usepackage{subfigure}
\usepackage{array}

\usepackage{epsfig}
\usepackage{natbib}
\usepackage{rotating}

\shorttitle{Dynamically Driven Inflow}
\shortauthors{Hatchfield et al.}
\begin{document}

\title{Dynamically Driven Inflow onto the Galactic Center and its Effect upon Molecular Clouds}

\correspondingauthor{H Perry Hatchfield}
\email{h.hatchfield@uconn.edu}

\author{H Perry Hatchfield}
\affiliation{University of Connecticut, Department of Physics, 196A Auditorium Drive, Unit 3046, Storrs, CT 06269 USA}
\author{Mattia C. Sormani}
\affiliation{Universit\"at Heidelberg, Zentrum f\"ur Astronomie, Institut f\"ur Theoretische Astrophysik, Albert-Ueberle-Str. 2, D-69120 Heidelberg, Germany}
\author{Robin G. Tress}
\affiliation{Universit\"at Heidelberg, Zentrum f\"ur Astronomie, Institut f\"ur Theoretische Astrophysik, Albert-Ueberle-Str. 2, D-69120 Heidelberg, Germany}
\author{Cara Battersby}
\affiliation{University of Connecticut, Department of Physics, 196A Auditorium Drive, Unit 3046, Storrs, CT 06269 USA}
\author{Rowan J. Smith}
\affiliation{Jodrell Bank Centre for Astrophysics, Department of Physics and Astronomy, University of Manchester, Oxford Road, Manchester M13 9PL, UK}
\author{Simon C.O. Glover}
\affiliation{Universit\"at Heidelberg, Zentrum f\"ur Astronomie, Institut f\"ur Theoretische Astrophysik, Albert-Ueberle-Str. 2, D-69120 Heidelberg, Germany}
\author{Ralf S. Klessen}
\affiliation{Universit\"at Heidelberg, Zentrum f\"ur Astronomie, Institut f\"ur Theoretische Astrophysik, Albert-Ueberle-Str. 2, D-69120 Heidelberg, Germany}
\affiliation{Universit\"{a}t Heidelberg, Interdisziplin\"{a}res Zentrum f\"{u}r Wissenschaftliches Rechnen, Im Neuenheimer Feld 205, D-69120 Heidelberg, Germany}

\begin{abstract}
The Galactic bar plays a critical role in the evolution of the Milky Way's Central Molecular Zone (CMZ), driving mass inward toward the Galactic Center via gas flows known as dust lanes. To explore the interaction between the CMZ and the dust lanes, we run hydrodynamic simulations in {\sc arepo}, modeling the potential of the Milky Way's bar in the absence of gas self-gravity and star formation physics, and we study the flows of mass using Monte Carlo tracer particles. We estimate the efficiency of the inflow via the dust lanes, finding that only about a third (30$\pm$12\%) of the dust lanes' mass initially accretes onto the CMZ, while the rest overshoots and accretes later. Given observational estimates of the amount of gas within the Milky Way's dust lanes, this suggests that the true total inflow rate onto the CMZ is 0.8$\pm 0.6$ M$_\odot$~yr$^{-1}$. Clouds in this simulated CMZ have sudden peaks in their average density near the apocenter, where they undergo violent collisions with inflowing material. While these clouds tend to counter-rotate due to shear, co-rotating clouds occasionally occur due to the injection of momentum from collisions with inflowing material ($\sim 52\%$ are strongly counter-rotating, and $\sim 7\%$ are strongly co-rotating of the 44 cloud sample). We investigate the formation and evolution of these clouds, finding that they are fed by many discrete inflow events, providing a consistent source of gas to CMZ clouds even as they collapse and form stars. 


\end{abstract}

\section{Introduction}
\label{sec:introduction}
The process of star formation is governed by a complex interweaving of physical mechanisms including self-gravity, galactic scale dynamical effects, turbulence, stellar feedback and magnetic fields \citep{mckee_Theory_2007, federrath_inefficient_2015, klessen_Physical_2016}. Studying the individual effects and interaction of these mechanisms and how they vary as a function of environment is necessary to understand star formation as it occurs across the cosmos. 

A vast range of physical scales are relevant to the processes that govern star formation, ranging from the kiloparsec scales typical of galactic features such as spiral arms and bars, down to the $< 0.01$ parsec scales of individual protostellar systems. Hydrodynamic simulations have proven to be an excellent tool for studying the interplay of gas self-gravity, turbulence (e.g.\ \citealt{federrath_Star_2012}), galactic dynamics (e.g.\ \citealt{kim_central_2012, armillotta_Life_2019,  kruijssen_dynamical_2019, tress_Simulations_2020a, smith_Cloud_2020,Li+2021}), star formation and feedback processes such as supernovae and stellar winds across this range of scales (e.g.\ \citealt{federrath_Modeling_2014, walch_SILCC_2015, girichidis_SILCC_2016, gatto_SILCC_2017, rahner_warpfield_2019, rosen_role_2020, tress_Simulations_2020b, sormani_Simulations_2020}). Observationally, however, the smallest of these scales can only be studied within the Milky Way. Since there is a limited variety of conditions within our Galaxy, this provides only a small window into the environmental variation of star formation as it occurs across the cosmos.

The Milky Way's center provides a unique opportunity to study the physics of star formation on sub-parsec scales within a more extreme environment than those nearby within the Milky Way's disk. The Central Molecular Zone (CMZ), which is defined as the region within Galactocentric radius $R\approx 250$~pc, hosts a reservoir of $\sim$ 3-5$\times$10$^7$~M$_\odot$ in molecular gas \citep{morris_GALACTIC_1996, dahmen_Molecular_1998}, with its clouds exhibiting high average densities and gas temperatures relative to the Galactic Disk (e.g.\ \citealt{mills_detection_2013, ginsburg_dense_2016, krieger_Survey_2017, mills_Dense_2018}), large velocity dispersions (e.g.\ \citealt{henshaw_molecular_2016, federrath_link_2016, kauffmann_galactic_2017, henshaw_Brick_2019}), and intense magnetic fields (e.g.\ \citealt{crutcher_zeeman_1996, pillai_Magnetic_2015}). These conditions bear similarities to some high redshift ($z\sim 2$) galaxies in which such detailed observations are impossible \citep{kruijssen_Comparing_2013}. 

The dynamics of the galactic environment are understood to be vital for modeling the mass flows, star formation rates and efficiencies, and molecular cloud life cycles within galaxy centers \citep[e.g.\ ][]{sormani_geometry_2019, tress_Simulations_2020b, sormani_Simulations_2020, orr_fiery_2021a}. In particular, tidal compression effects during pericenter passage have been suggested as a mechanism for triggering star formation in molecular clouds on CMZ orbits \citep{kruijssen_dynamical_2015,  jeffreson_physical_2018, kruijssen_dynamical_2019, dale_dynamical_2019}.
 
The Milky Way is known to host a stellar bar (e.g.\ \citealt{blitz_direct_1991, binney_photometric_1997,wegg_mapping_2013}). This bar generates a highly non-axisymmetric gravitational field within the Galaxy's inner few kpc, which induces strongly non-circular gas motions \citep{fux_3d_1999, kim_NUCLEAR_2011, kim_central_2012, li_hydrodynamical_2015, sormani_Gas_2015b, ridley_nuclear_2017, sormani_Dynamical_2018}. The bar potential admits two families of closed orbits that are of particular relevance for the gas flows: the x$_1$ orbits, which occur in the outer parts, have the semimajor axis aligned with the bar and become more eccentric as the Galactic Center is approached; and the mildly elliptical x$_2$ orbits, which occur closer to the Center and have semimajor axis perpendicular to the bar \citep{binney_Understanding_1991, sormani_Gas_2015a}. The self-interaction of gas on the innermost x$_1$ orbit induces shocks that efficiently drive material from the x$_1$ to the x$_2$ orbits, where the majority of CMZ gas settles \citep{sormani_Gas_2015a}. The shocked streams of gas transitioning between these orbital families are known as the bar "dust lanes" by analogy with similar structures observed in many barred spiral galaxies (such as NGC 1097, NGC 1300, and many others, see for example \ \citealt{Martini_circumnuclear_2003a, martini_Circumnuclear_2003}). 
Position-velocity features interpreted as the dust lanes have been identified in the CO maps of the Galactic Center \citep[e.g.\ ][]{fux_3d_1999,marshall_large_2008}, and are believed to be the primary avenue by which the Milky Way's CMZ gains mass (e.g.\ \citealt{kim_central_2012,tress_Simulations_2020b}). 

To measure this bar-driven inflow rate, \cite{sormani_mass_2019} estimate the mass and inwards velocity of material in the dust lanes by interpreting $^{12}$CO observations of longitude-velocity (l-v) structures near the Galactic Center using a simple geometrical model that is motivated by the hydrodynamical simulations from \citet{sormani_theoretical_2018}. The $^{12}$CO intensity from the structures that they identify as dust lane candidates is translated to a mass using an X$_{\rm CO}$ conversion factor, and then the gas radial velocity is calculated using the geometrical model and adopting recent constraints on the bar's position-angle relative to the Sun \citep{wegg_mapping_2013, bland-hawthorn_galaxy_2016}. With this model, they find a mass inflow rate of $2.7^{+1.5}_{-1.7}$~M$_\odot$yr$^{-1}$. However, this estimate assumes that all inflowing mass in the dust lanes will accrete onto the CMZ on the timescale suggested by its inward velocity (i.e., its distance from the CMZ divided by its radial speed). They acknowledge (see their Section 4.1.4) that a significant fraction of the dust lane material likely overshoots the CMZ and will be added to the CMZ's mass on a longer timescale, and that as a consequence their simple model consequently tends to overestimate the inflow rate.

In this work, we run a hydrodynamic simulation of the flows of gas in the presence of the Milky Way's barred gravitational potential in order to measure the inflow efficiency of the dust lanes. We also use these simulations to study how the properties of molecular clouds are affected by the dynamics of the Galactic Center. The key novelty of this simulation is that we include tracer particles that are weighted by mass throughout the entire simulation. This allows us to study in detail the mass flows, and to keep track of the properties of molecular clouds as a function of time and reconstruct their history. We do not include gas self-gravity, star formation or stellar feedback to better isolate the effect of the external gravitational field and orbital properties on cloud properties.

The structure of this paper is as follows. In Section \ref{sec:design} we summarize the numerical setup for our simulated CMZ. Section \ref{sec:inflow} describes our tracer particle method for following material falling into the CMZ, and presents the results from our measurement of the inflow efficiency. In Section \ref{sec:clouds}, we use tracer particles to follow the evolution, angular momenta, and build-up of individual clouds in the CMZ, and we report the results of this cloud-tracing analysis. We discuss the possible implications of these results in Section \ref{sec:discussion}. Finally, we conclude and sum up in Section \ref{sec:summary}.

\section{Simulation Design}
\label{sec:design}
We use the moving-mesh code {\sc arepo} \citep{springel_pur_2010,weinberger_Arepo_2020} to re-simulate the model of \cite{sormani_geometry_2019} with the addition of Monte Carlo tracer particles (MCTP, or just tracer particles). These allow us to follow the flow of mass between cells and track the properties of clouds self-consistently between simulation snapshots. Here we briefly summarize the numerical design of these simulations. A more detailed description of the simulation design can be found in \cite{sormani_theoretical_2018} and \cite{sormani_geometry_2019}, while a description of the implementation of the MCTP can be found in \citet{genel_Following_2013}. 

\begin{figure*}
\begin{center}
\includegraphics[trim = 0mm 0mm 0mm 0mm, width = .85 \textwidth]{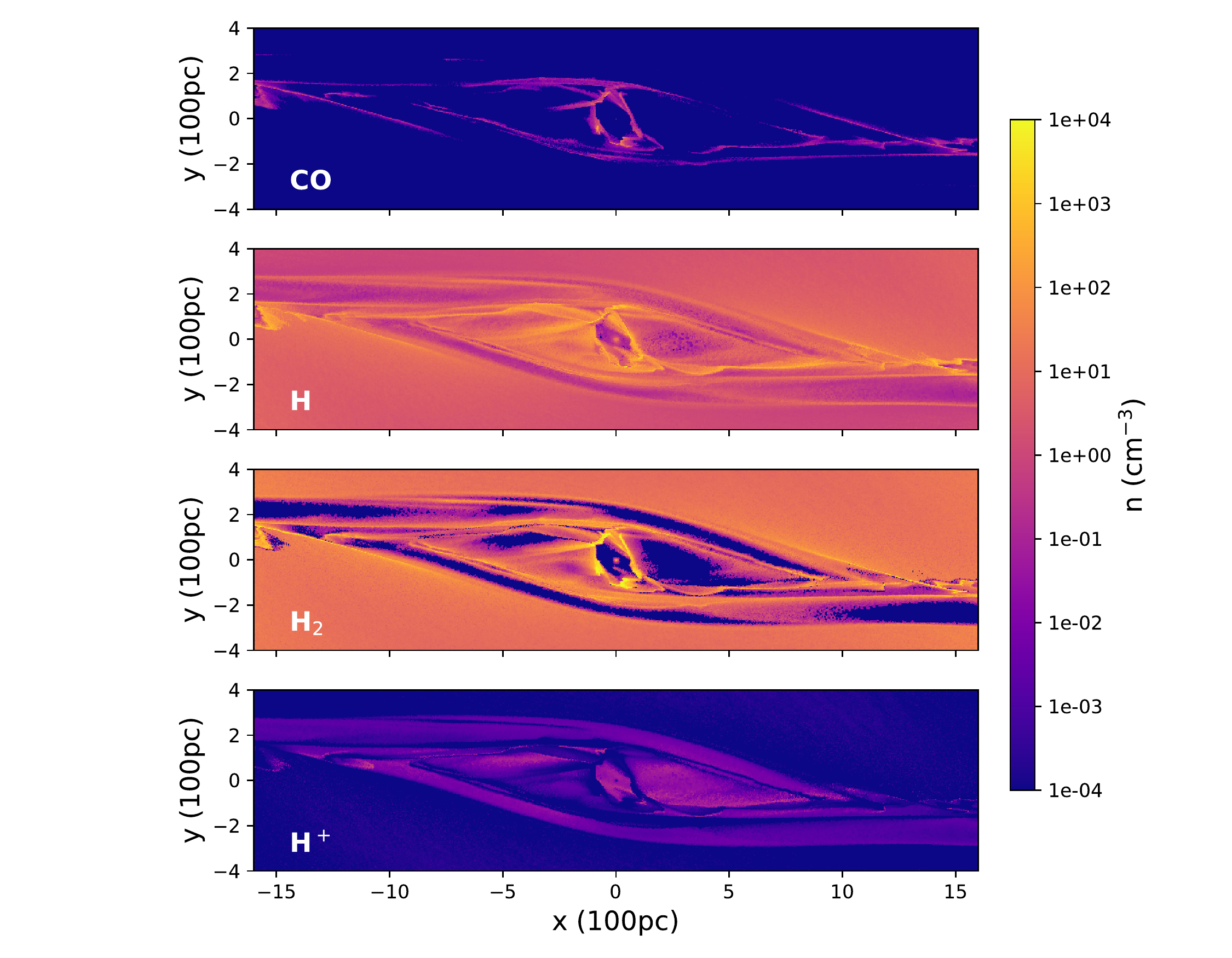}
\end{center}
\caption{Face-on density slices through the plane of $z=0$ for the distribution of four chemical species traced by our chemical network. From top to bottom CO, H$_2$, H and H$^+$ are shown. The gas in the Galactic Center is highly molecular, with high densities mainly concentrated in the dust lanes and in the central x$_2$ orbits.}
\label{fig:chem_snap_gallery}
\end{figure*}

{\sc arepo} solves the equations of hydrodynamics in three-dimensions in the absence of magnetic fields\footnote{{\sc arepo} can also be used to model magnetohydrodynamical flows, but we do not make use of these capabilities in this study.} and gas self-gravity:
\begin{equation}
	\frac{\partial \rho}{\partial t} + \nabla \cdot (\rho \mathbf{v}) = 0
\end{equation}
\begin{equation}
	\frac{\partial \rho \mathbf{v}}{\partial t} + \nabla \cdot (\rho \mathbf{v} \otimes \mathbf{v} + P \mathbf{I}) \mathbf{v} = -\rho \nabla \Phi
\end{equation}
\begin{equation}
	\frac{\partial (\rho e)}{\partial t} + \nabla \cdot [(\rho e + P)\mathbf{v}] = \dot{Q} + \rho \frac{\partial \Phi}{\partial t}
\end{equation}
where $\rho$ is the density of the gas parcel, $\mathbf{v}$ is the fluid velocity, $P$ is the thermal pressure field, $\mathbf{I}$ is the identity matrix, $\boldsymbol\Phi$ is the external gravitational potential tuned to fit the properties of the Milky Way, $e=e_{\rm thermal} + \boldsymbol\Phi + \mathbf{v}^2/2$ is the total energy per unit mass and $e_{\rm thermal}$, the thermal energy per unit mass, and $\dot{Q}$ describes the rate of change in the internal energy of the cell due to radiative and chemical cooling and heating. We assume an ideal gas equation of state, obeying 
\begin{equation}
P = (\gamma-1)\rho e_{\rm thermal}
\end{equation}
with an adiabatic index $\gamma=5/3$, a good approximation for both the atomic and cold molecular hydrogen composing the vast majority of the gas content in our regions of interest \citep{sormani_theoretical_2018}. 
We use the NL97 chemical network from \cite{Glover_approximations_2012} to follow the interaction and evolution of chemical species in our simulated ISM. This network combines a model for hydrogen chemistry from \citet{glover_Simulating_2007} with a simple model for CO formation and destruction developed by \cite{Nelson_dynamics_1997}. It tracks the non-equilibrium abundances of the species H, H$_2$, H$^+$, C$^+$, O, and CO between each timestep. The distributions of several species is shown for a sample timestep in Figure \ref{fig:chem_snap_gallery}.

We initialize the simulation with an initial axisymmetric density profile described by 
\begin{equation}
\rho_{\rm init}(R,z) = \frac{\Sigma_0}{4z_d}\text{exp}\bigg(-\frac{R_m}{R}-\frac{R}{R_d}\bigg) \text{sech}^2 \bigg(\frac{z}{2z_d}\bigg),
\end{equation}
for Galactocentric radius $R$ and height $z$, $\Sigma_0=50$~M$_\odot$pc$^{-2}$, $z_d=85$~pc, $R_d=7$~kpc, and $R_m=1.5$~kpc. These parameters have been chosen to match the observed radial distribution of gas in the Milky Way \citep{Kalberla_global_2008, Heyer_molecular_2015}. In order to cut down on computational expense, the outer reaches of the Galactic Disk (all material at $R\geq 10$~kpc) are removed from the simulation's initial conditions. Gas beyond this radius has a negligible effect on the gas dynamics at smaller Galactic radii over the relevant timescale. No noise is added to the initial distribution of gas, and all asymmetries and turbulent flows observed in the gas flows are caused by the imposed barred Galactic potential. 

In this work we use the same rigidly rotating external barred potential as used by \citet{sormani_geometry_2019}. This potential is composed of components for the halo, disk, bulge, and a bar rigidly rotating with pattern speed $\Omega_p=40$~km s$^{-1}$kpc$^{-1}$ whose properties are consistent with observational constraints \citep{Launhardt_nuclear_2002, sormani_Gas_2015b, mcmillan_mass_2017, portail_dynamical_2017, sanders_pattern_2019, sormani_Simulations_2020, bland-hawthorn_galaxy_2016}. For more details on the explicit form and construction of the gravitational potential, we refer the reader to appendix A of \cite{sormani_geometry_2019}. The non-axisymmetric potential term, corresponding to the stellar bar, is introduced slowly from the beginning of the simulated time in order to avoid transient dynamics in the early gas distribution. The barred potential is fully developed at $t\approx150$~Myr, and all of the analysis is performed on simulation snapshots later than this time. The simulation is run for $\sim$30 Myr after the barred potential is fully active.

\begin{figure*}
\begin{center}
\includegraphics[trim = 0mm 0mm 0mm 0mm, width = .9 \textwidth]{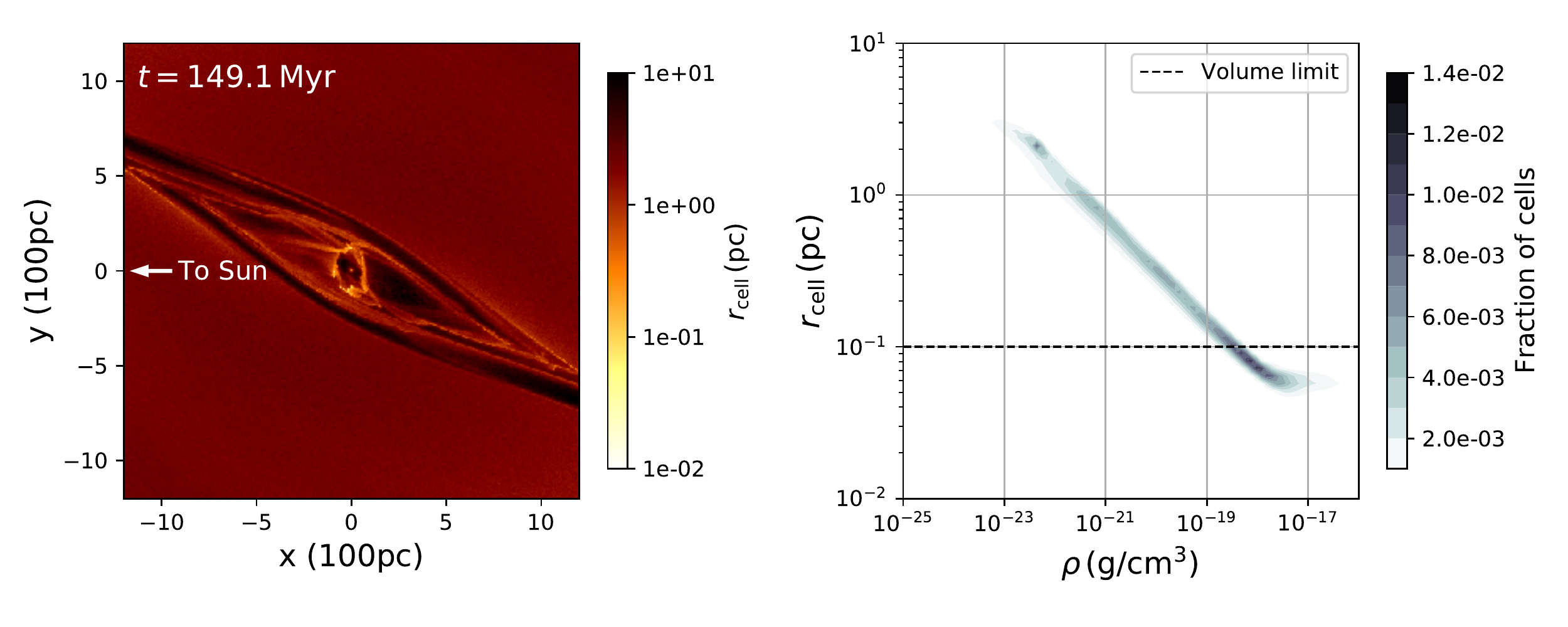}
\end{center}
\caption{Visualizations of the distribution of cell spatial resolution across the simulation. The left-hand panel shows cell radius from top down slice through $z=0$, showing how the spatial resolution is higher for cells in regions of interest such as the dust lanes and x$_2$ orbits in the CMZ. The snapshot has been rotated such that the sun's position is directly to the left. The right-hand panel shows the distribution of cell radius as a function of density, highlighting the volume cutoff corresponding to an effective radius of 0.1~pc to prevent runaway cell division and computational slowdown. Cells with volumes smaller than half this cutoff are derefined in the next timestep, and are not permitted to sub-divide further. The right-hand panel indicates that resolution scale is largely a proxy for cell density. The bar has been rotated 20$^\circ$ relative to other figures to include more of the dust lane region in the frame.}
\label{fig:refinement}
\end{figure*}

The spatial resolution of our simulation varies depending on the local density. Cells are refined down to a target cell mass of $25$~M$_\odot$. To achieve this, {\sc arepo} divides any cells with mass a factor of two larger than the target mass, and merges any adjacent cells that are a factor of two less massive so that cells are kept on average close to the target mass. We also implement a minimum effective cell radius ($r_{\rm eff} = ({3 V_{\rm cell}}/{4\pi})^{1/3}$ for cell volume $V_{\rm cell}$) of $0.1$~pc to prevent excessive refinement and computational slow-down in areas of extremely high density. Cells with a volume below this cell volume are not permitted to divide into smaller cells. This refinement scheme allows for a higher spatial resolution in regions of high density while lowering the computational cost of including large portions of lower density material outside of the Galactic center region. The distribution of cell spatial resolution is shown in Figure \ref{fig:refinement}, which shows that the densest gas is resolved with cells of radius $\sim0.1$~pc. 

To trace the mass flows in this simulation, we use the Monte Carlo tracer particle implementation presented in \cite{genel_Following_2013}. The tracers are massless particles associated, at any given timestep, with a given {\sc arepo} cell, known as that tracer's ``parent cell''. These tracers do not have individual positions or velocities of their own, instead adopting those of their parent cell. Using the instantaneous finite volume fluxes, which are already calculated for each Voronoi cell face between every active timestep, we determine a mass-flux weighted number of tracers to be transferred across each cell interface. This allows the tracers to effectively follow the fluid flow. The Monte Carlo draw occurs in order to probabilistically distribute the finite number of tracers in each cell according to the mass fluxes. Without the tracer particles, the mass flow cannot be easily followed between multiple timesteps because {\sc arepo} is not a purely Lagrangian code. More details about the implementation of this procedure and its advantages over other tracer particle schemes can be found in \cite{genel_Following_2013}. Tracers are injected into each cell in the simulation in a mass-weighted fashion at a simulation time of 100 Myr, with roughly 1 tracer per 10 M$_\odot$, rounded up. Given our adopted target mass, this means that we have roughly 2-3 tracers per simulation cell, with more massive cells having more tracers. These tracers then move according to the mass fluxes for the remaining simulated time.

\begin{figure}
\begin{center}
\includegraphics[trim = 25mm 0mm 0mm 0mm, width = .49 \textwidth]{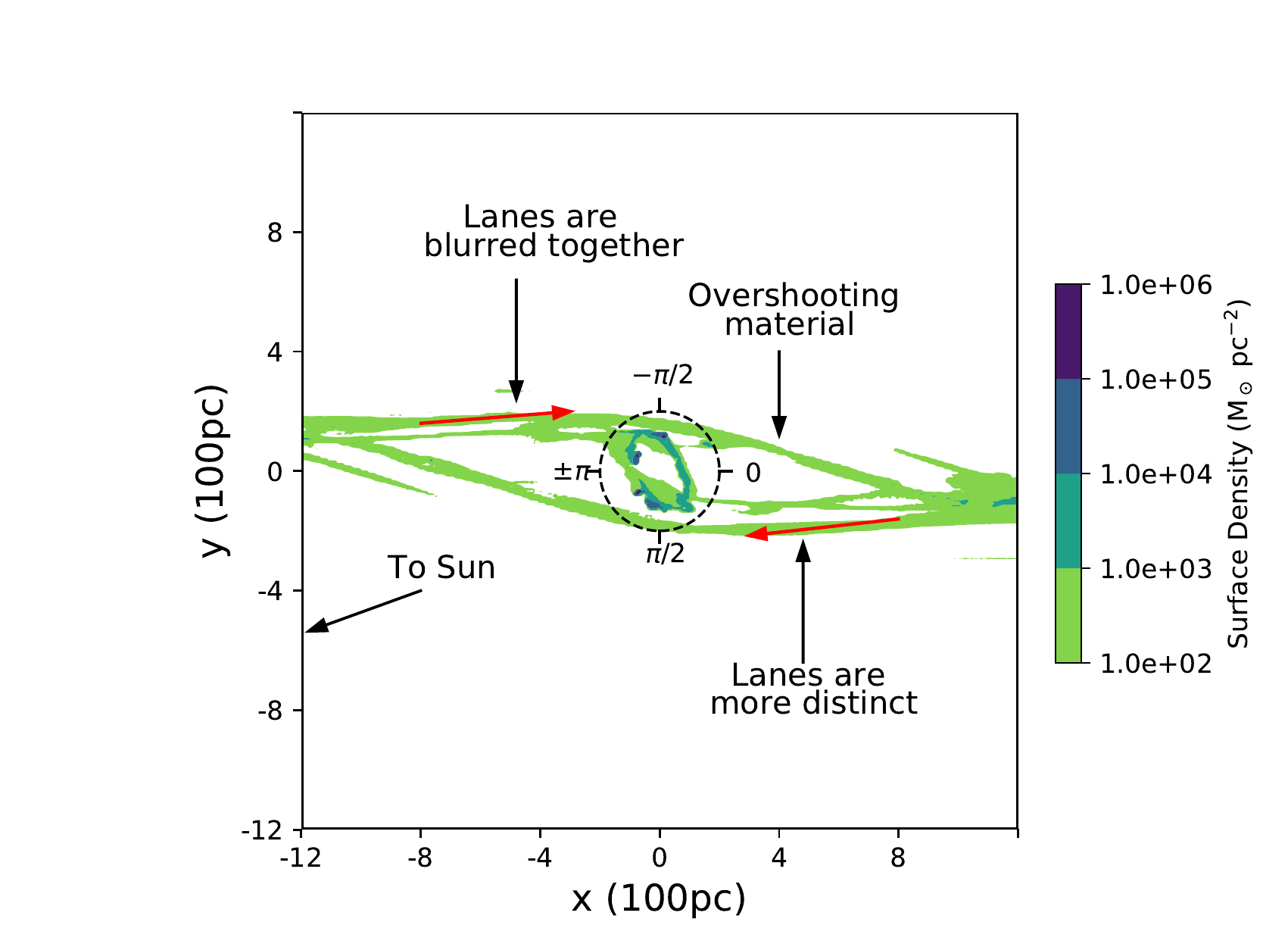}
\end{center}
\caption{A schematic illustration of the simulated gas flow features at simulation time $t=149$ Myr, where the bar major axis has been oriented horizontally. The dotted circle indicates the CMZ, with phase angles labeled according to the notation used in this work. The red arrows denote the flow of gas along the near-side and far-side dust lanes. An example of the blurring of the dust lane components is highlighted in the upper left-hand side, while a more distinct configuration is highlighted on the lower right-hand side. The direction to the observer (the Sun) is indicated.}
\label{fig:cartoon}
\end{figure}

\section{The Inflow Efficiency of Dust Lane Gas Accreting into the CMZ}
\label{sec:inflow}

While the Galactic bar is an efficient mechanism for driving gas to smaller Galactocentric radii, not all of the gas in the dust lanes accretes directly into the innermost 250~pc of the Galaxy on the timescale naively implied by its inwards velocity. Some of the high-velocity gas transported by the dust lane shocks will overshoot, and be accreted at a later time. This inflow inefficiency has been noted in the past.
For example, the results of the hydrodynamical model of the barred galaxy NGC 1530 presented in \citet{Regan_mass_1997} show that not all of the material traveling along that galaxy's dust lanes is efficiently accreted into the CMZ, and some instead overshoots and interacts with the opposite dust lane. However, the precise fraction of overshooting gas calculated by \citet{Regan_mass_1997} was inaccurate due to an inherited mistake in the treatment of the bar using the CMHOG code \citep{kim_central_2012}. This overshooting phenomenon is also visible by eye in previous simulations of the Milky Way's Galactic Center as well (e.g.\ \citealt{sormani_theoretical_2018}), but could not be quantified because individual mass elements cannot be followed within {\sc arepo}'s quasi-Lagrangian framework. In this section, we measure the inflow efficiency of inflowing gas by following samples of tracer particles selected from the dust lanes. 

\subsection{The Morphology of the Dust Lanes}
Due to the complexity and variability of the dust lanes, it is not immediately obvious how we should define the inflow efficiency. To do so effectively, we must consider the morphology of the dust lanes and how matter accretes onto the CMZ. In the simulations presented in this work, the dust lanes appear to generally have multiple components that span a range in velocity and impact parameter with respect to the Galactic Center. Multiple, spatially distinct dust lanes are not a general feature of simulations of gas flow in barred galaxies, but similar velocity features that resemble these components are visible in position-velocity maps of CO within the Milky Way's Galactic Center (see the features labeled L1 and L3 in figure 1 and their simulated counterparts D1 and D2 in figure 3 of \citealt{sormani_geometry_2019}; see also \citealt{Liszt2008}). The specific morphology of the simulated dust lanes depends on the details of the gravitational potential used in the simulation and the sub-grid feedback physics, and additionally varies as a function of simulation time. A simplified schematic of a sample dust lane morphology and a general description of the dust lane region and CMZ that emerges from these simulations is shown in Figure~\ref{fig:cartoon}.

The simulations of \citet{tress_Simulations_2020b}, which are identical to the simulations presented here except for the inclusion of gas self-gravity and supernova feedback (and in particular utilize the same external gravitational potential and initial gas distribution) show that supernova feedback can spatially blur together the two dust lanes. They also show that the total mass inflow rate toward the CMZ from the dust lanes is not dependent on the inclusion of gas self-gravity or star formation feedback. This suggests that the blurring of the dust lanes does not affect the average inflow rate, which is ultimately controlled by the torques of the bar potential as they remove angular momentum from the gas. Therefore, it should not affect the average correction factor that we aim to derive in this work.

Given this understanding of the dust lane morphology and of the inflow mechanism, we measure the inflow efficiency of our simulated Galactic Center using three different methods. The first method compares the rate derived from an approach resembling that of \cite{sormani_mass_2019} to the actual CMZ growth rate. In the second method, we measure the probability that a randomly selected parcel of mass in the dust lane will be accreted the first time it passes the CMZ. In the final method, we constrain the number of times infalling tracers pass within 250 pc of the Galactic Center without remaining within the CMZ indefinitely (or whether they never get accreted). As we will see below, the values of the inflow efficiency obtained by these three different methods agree with each other within their uncertainties.

\subsection{Measuring the Bulk Inflow Efficiency of the Dust Lanes}
\label{sec:inflow_bulk}

Firstly, we measure the true inflow rate in the simulations by measuring the mass growth of the CMZ as a function of time. The total gas mass of the CMZ is obtained by summing up the mass of all cells with Galactocentric radius $<250$ pc as a function of time, and the inflow rate is calculated as the rate of change of this total mass. We find that while the instantaneous inflow rate can vary within a factor of $\sim 2$, the average value over a 30 Myr window is $\sim$0.7 M$_\odot$ yr$^{-1}$. 

We then calculate a ``naive estimate'' of the inflow rate by assuming that all of the inflowing mass in the dust lanes will accrete onto the CMZ within the time suggested by its radially inward velocity. In other words, for each cell in the dust lane region that has an inward radial velocity greater than 50 km s$^{-1}$, we consider its projected velocity toward the Galactic Center and its mass, and treat each of these parcels of mass as if they will accrete within a time $t=R_{\text{GC}}/v_{\text{in}}$ where $R_{\text{GC}}$ is the Galactocentric Radius of the cell and $v_{\text{in}}$ is the radial component of the velocity toward the Galactic Center. However, this method is likely to overestimate the accretion rate because it assumes that all the gas accretes the first time that it passes close to the CMZ, while in reality it will sometime overshoot as noted above. Considering the three independent dust lane samples selected throughout the simulation, we obtain estimates of the inflow rate ranging between 1.5-1.8 M$_\odot$ yr$^{-1}$ with a mean of 1.7 M$_\odot$ yr$^{-1}$. 

Therefore, taking the ratio of the true net mass inflow rate of $\sim$0.7 M$_\odot$ yr$^{-1}$ to the naive estimate of $\sim$1.7 M$_\odot$ yr$^{-1}$, we obtain an inflow efficiency of $\epsilon\approx 0.41$. This value fluctuates on smaller ($\sim$1 Myr) timescales, but these variations are averaged out over the timescale used in this calculation (30 Myr). The other methods that are described below will give an indication of the small timescale variations.

\subsection{Measuring the Instantaneous Inflow Efficiency of the Dust Lanes }\label{sec:inflow_inst}

A second way to measure the inflow efficiency is to consider the probability that any given parcel of mass in the dust lanes will be accreted onto the CMZ on its next pass within 250 pc of the Galactic Center. This approach also allows us to better estimate the variability of the inflow efficiency.

\begin{figure}
\begin{center}
\includegraphics[trim = 20mm 0mm 0mm 0mm, width=0.45 \textwidth]{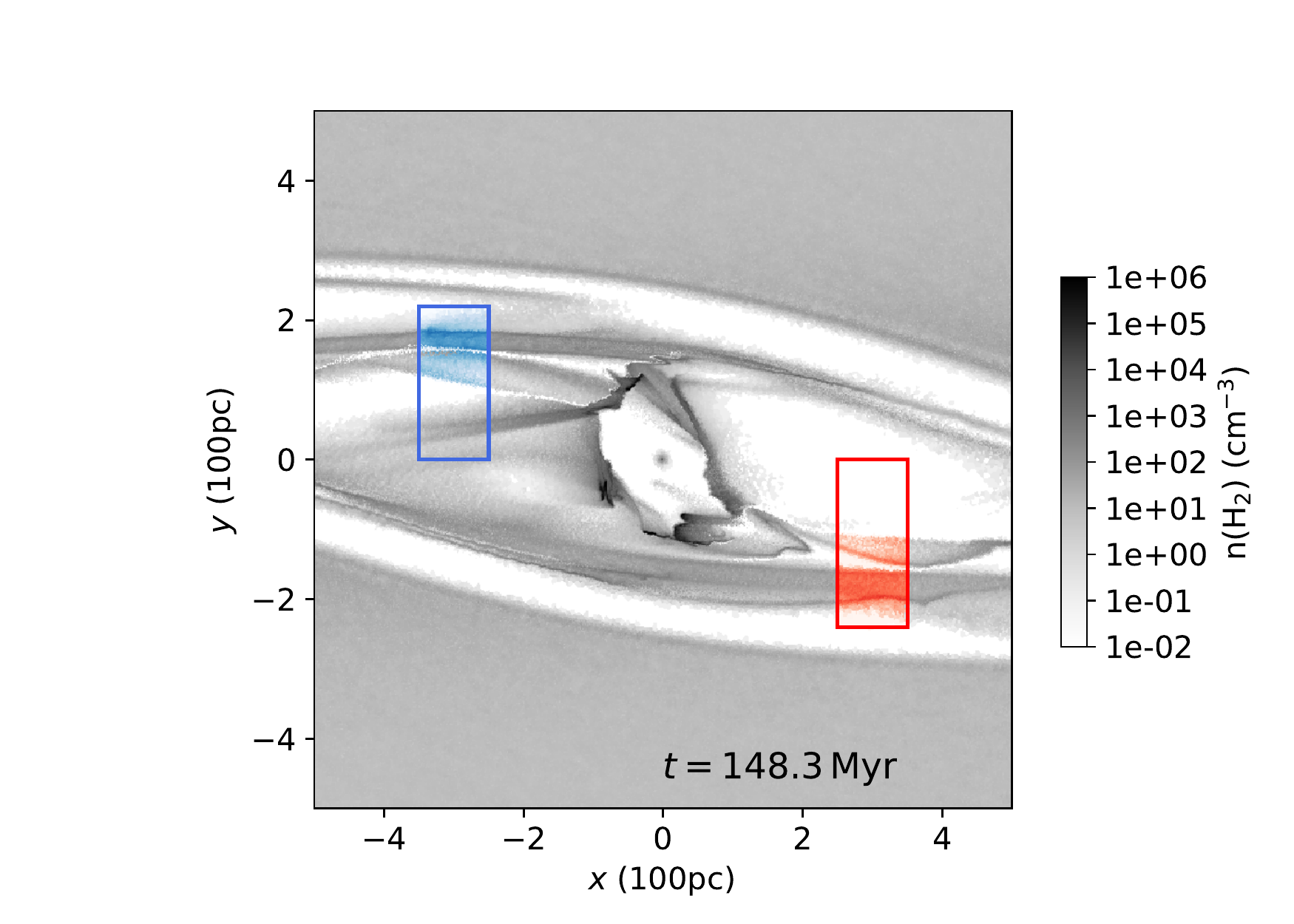}
\end{center}
\caption{An example of the selection of inflowing material used to measure the inflow efficiency in Section \ref{sec:inflow_inst}. The tracer particles are selected from all cells within 100~pc long segments of the near and far-side dust lanes (shown as blue and red rectangles, respectively) that meet a velocity criteria ($v<50$~km s$^{-1}$) to exclude cells that are not associated with the dust lanes. The cells hosting selected tracers are colored blue and red corresponding to the near and far-side dust lanes, and the corresponding colors in Figure \ref{fig:inflow_percent}. Modest changes to the shape, orientation, and length of the selection regions do not substantially change the inflow efficiency measured. To calculate $\langle\epsilon_{\rm DL}\rangle$ in Section \ref{sec:inflow_inst}, 25 pairs of tracer regions (for a total of 50 samples) are selected from different timesteps throughout the simulation. The blue selection of tracers is used as an example in Section \ref{sec:inflow_history}.}
\label{fig:inflow_setup}
\end{figure}

\begin{figure*}
\begin{center}
\includegraphics[trim = 0mm 0mm 0mm 0mm, width = .98 \textwidth]{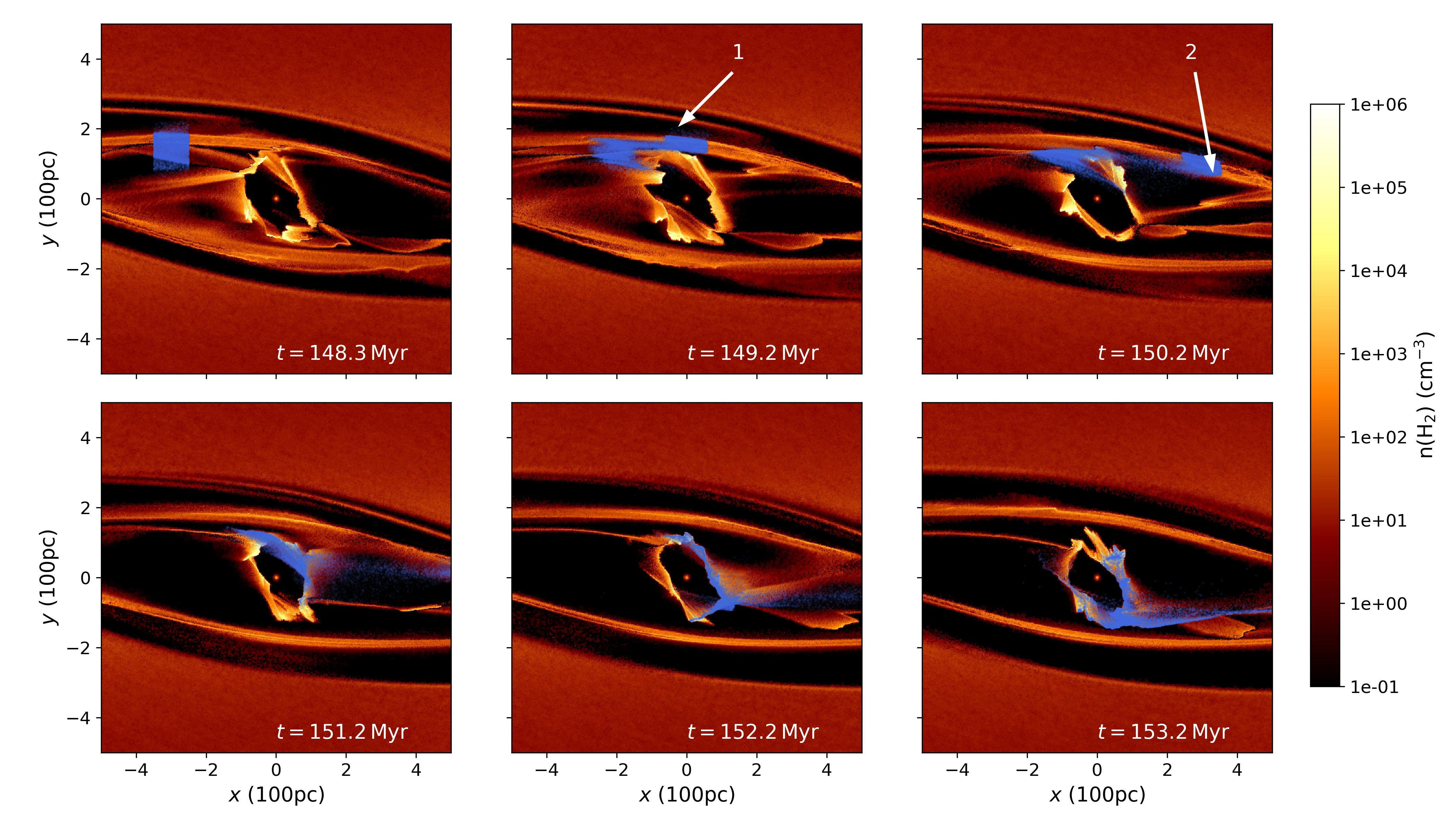}
\end{center}
\caption{Six snapshots depicting the tracer inflow over time for one of the 50 tracer selections, plotted in blue over that timestep's molecular hydrogen number density. The snapshot is rotated such that the Sun's position is directly to the left, as in Figure \ref{fig:refinement}. The blue tracers correspond to the same selection used to generate the left-hand panel of Figure \ref{fig:inflow_percent}. In this case the tracers are selected from the near-side dust lane, in which the two velocity components are blurred together. The top middle panel shows the material on its initial infalling trajectory, labeled 1. In the top right-hand panel, the overshooting material is labeled 2, and the rest of the material is undergoing a violent collision as it accretes onto the CMZ.}
\label{fig:inflow_snap_gallery}
\end{figure*}

\begin{figure*}
\begin{center}
\includegraphics[trim = 0mm 0mm 0mm 0mm, width = .9 \textwidth]{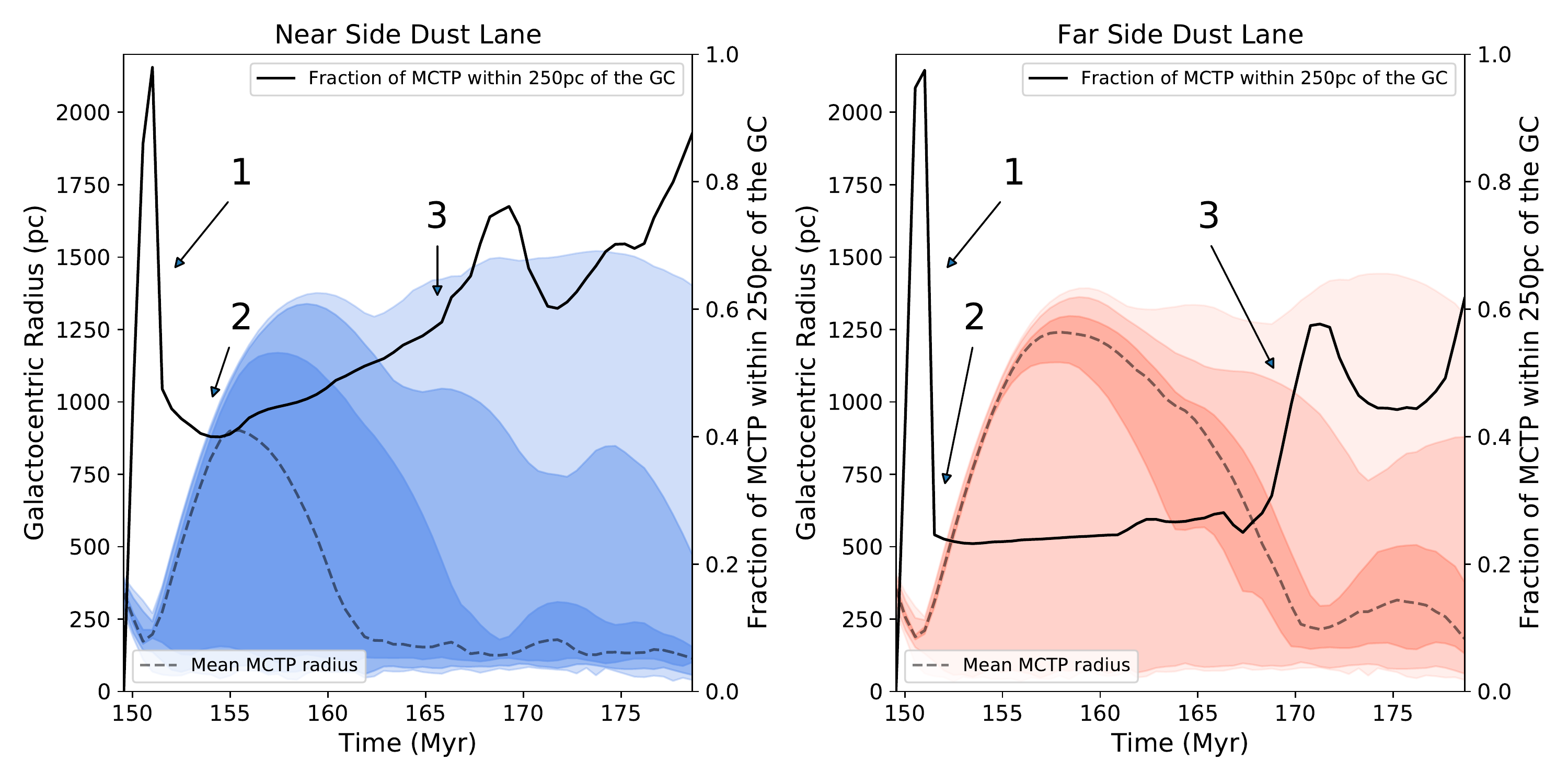}
\end{center}
\caption{The Galactocentric radial distribution of gas selected from infalling material in the dust lanes. The left-hand panel is the near-side (closer to the Sun) dust lane, and the right-hand panel shows the far-side dust lane. A dashed line shows the mean tracer particle radius at each timestep, and the solid black line is the percentage of tracer particles located within 250~pc of the Galactic Center. The blue and red contours show the radius containing the radius range that contains 33\%, 66\% and 100\% of the tracer particles both above and below the mean tracer value. In both panels, three points of interest are labeled, and correspond to similar labels in other figures (such as Fig. \ref{fig:inflow_snap_gallery}). Feature 1 shows the initial accretion event in which most of the gas passes within 250pc of the Galactic Center. Feature 2 highlights the dip in the percentage due to overshooting gas, and the inflow efficiency is measured as the local minimum around this point. Feature 3 designates the re-accretion as gas begins to fall in toward the CMZ along the opposite dust lane.}
\label{fig:inflow_percent}
\end{figure*}

We select 25 pairs of tracer particle samples (one from the near and one from from the far-side dust lane for each used timestep, for a total of 50 independent samples each containing $\sim10^5$ tracer particles), spaced $\sim$2.5 Myr apart to avoid excessive correlations. For each timestep used, the tracer particles are selected from a 100~pc long segment of the dust lanes starting at a radial distance of about 300~pc from the Galactic Center. An example of this selection for a single timestep is shown in Figure \ref{fig:inflow_setup}, and a gallery of snapshots following this same sample of tracers (the blue selection) as they overshoot and accrete is shown in Figure \ref{fig:inflow_snap_gallery}. Because the number of tracer particles injected into each cell is proportional to the cell mass, the sample effectively tracks the mass flows of the gas initially contained in the selected region. We ensure that the sample only contains material associated with the dust lanes by rejecting any tracers associated with cells having small ($v<50$~km s$^{-1}$) Galactocentric radial velocities. This velocity minimum also ensures that the 20 Myr window will be more than sufficient to observe the overshooting of the tracers because it would take a particle moving at $v>50$~km s$^{-1}$ less than 10 Myr to cross the 250 pc radius ring used to define our CMZ, and the vast majority of tracers in the dust lane are moving significantly faster than that. 
Using these tracer particle samples, we calculate the average instantaneous inflow efficiency $\langle \epsilon_{\rm DL}\rangle$ as
\begin{equation}
    \langle \epsilon_{\rm DL}\rangle = \bigg\langle \frac{\min(N_{\rm 250pc})_{20 \rm Myr}}{N_{\rm tot}} \bigg\rangle,
\end{equation}
where $\min(N_{\rm 250pc})_{20 \rm Myr}$ is the local minimum number of tracer particles contained within 250~pc of the Galactic Center ($N_{\rm 250pc}$) within a 20 Myr window after the tracers' first approach toward the CMZ, $N_{\rm tot}$ is the total number of tracers in the sample, and the angle brackets denote an average over the 50 samples of tracers. The time span of 20 Myr is only used to limit the time that we follow each tracer sample to cut down on unnecessary computational expense. The 20 Myr window that we used to select the minimum of $N_{\rm 250pc}$ begins only after the selected tracers undergo their initial approach to the CMZ. This measurement of the local minimum is visualized in Figure \ref{fig:inflow_percent}, which shows the percentage of tracer particles at a given Galactocentric radius. We see that the tracers initially approach the CMZ (at which time the 20 Myr window begins, labeled 1) before partially overshooting (labeled 2) and eventually re-accreting over time (labeled 3). The local minimum at which $\min(N_{\rm 250pc})_{20 \rm Myr}$ is calculated approximately corresponds to the point labeled 2.

We find, averaging the near and far-side dust lane, that $\langle \epsilon_{\rm DL} \rangle = 0.30\pm0.12$. Most of the gas that overshoots the CMZ collides with the opposite dust lane, and after $\sim$25 Myr, the majority of the overshooting mass of the initial parcel of gas has been accreted to within 250~pc of the Galactic Center. The uncertainty in the measurement of $\langle \epsilon_{\rm DL} \rangle$ reflects the standard deviation of the 50 tracer samples considered.

\subsection{Constraining the Inflow Efficiency using Individual Tracer Histories}
\label{sec:inflow_history}

Lastly, we can constrain the inflow efficiency by considering how many times a typical parcel of inflowing mass will overshoot before finally accreting onto the CMZ. We can provide an upper limit on the inflow efficiency by measuring the cumulative Galactocentric phase angle traveled by each tracer particle within the sample of inflowing tracer particles before they each finally accrete onto the CMZ. In this case, by ``finally accretes'', we mean the time at which each tracer arrives within 250 pc of the Galactic Center and stays within that radius for the remaining simulated time.

\begin{figure}
\begin{center}
\includegraphics[trim = 20mm 0mm 0mm 0mm, width = .48 \textwidth]{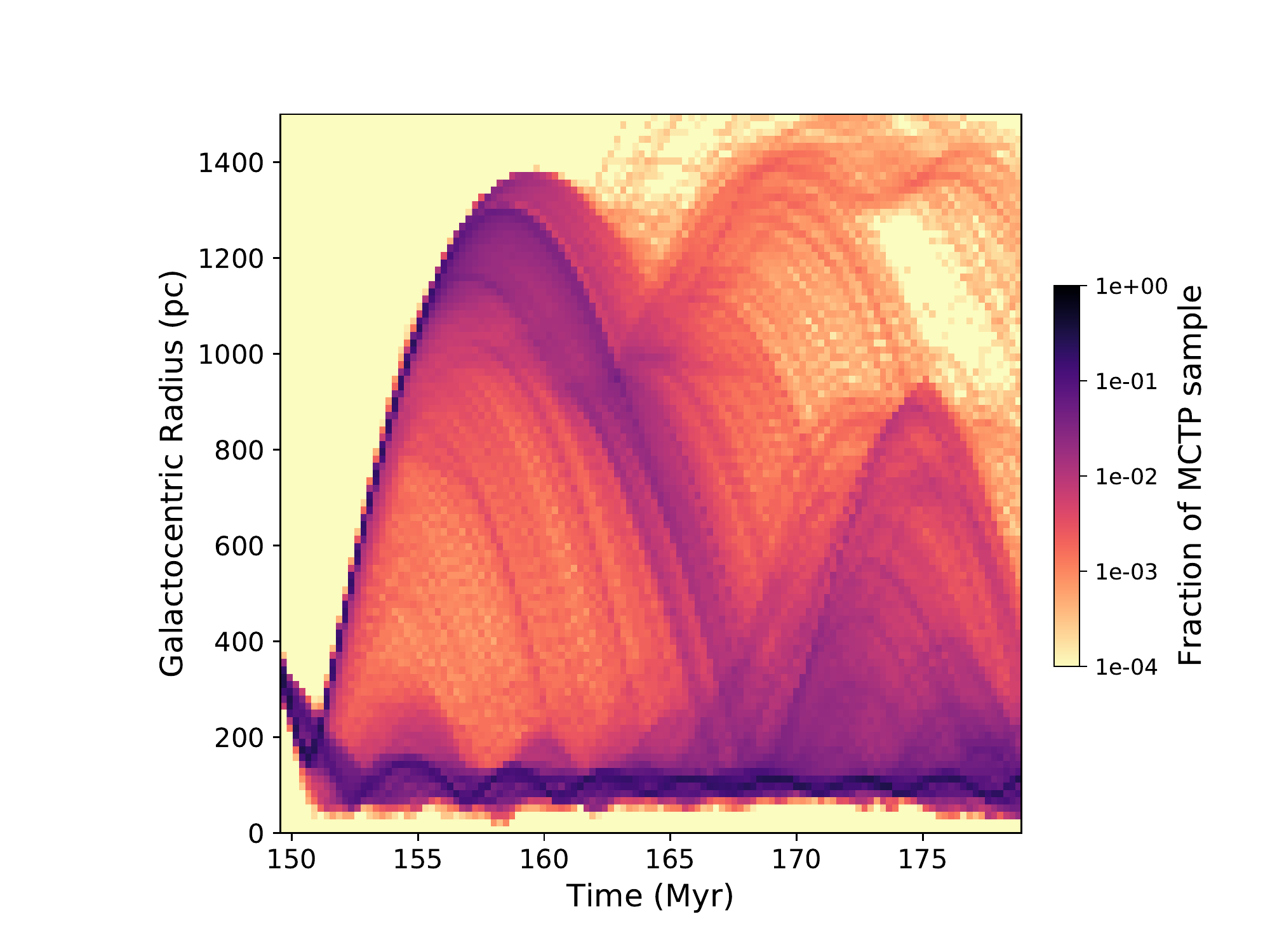}
\end{center}
\caption{The evolution of infalling tracer particles (the blue selection from Figure \ref{fig:inflow_setup}) as a function of simulation time and Galactocentric radius. Many of the tracer particles accrete onto the CMZ upon their first approach, while others overshoot, traveling to larger Galactocentric radii before either accreting or overshooting again. While the majority of the tracers are accreted within the 30 Myr time span simulated, about $\sim$23\% of the tracers do not accrete within the time span.}
\label{fig:inflow_gcrad}
\end{figure}

To perform this experiment, we consider a sample of tracers from the near-side dust lane. The sample used is identical to the blue sample in Figure \ref{fig:inflow_setup}. We follow the evolution of these tracers and identify the time at which they are ``finally accreted''. The evolution of this tracer particle sample is shown in Figure \ref{fig:inflow_gcrad}, in which the tracer particles are binned according to the simulation time and their Galactocentric radius. To quantify the ``final accretion'' time for each particle, we determine the simulation time at which the particle enters a Galactocentric radius of 250 pc and does not leave that radius for longer than 4 Myr. The window is chosen to be 4 Myr because >95\% of the mass in the dust lanes is moving with an inward velocity greater than 150 km s$^{-1}$. Moving constantly at that speed, it would take the tracers only $\sim3$Myr to cross the 500 pc diameter of the Galactic center. Employing this simple method, we can make the generalizing assumption that overshooting tracers will not be counted as accreted in this way. 

\begin{figure}
\begin{center}
\includegraphics[trim = 0mm 0mm 0mm 0mm, width = .45 \textwidth]{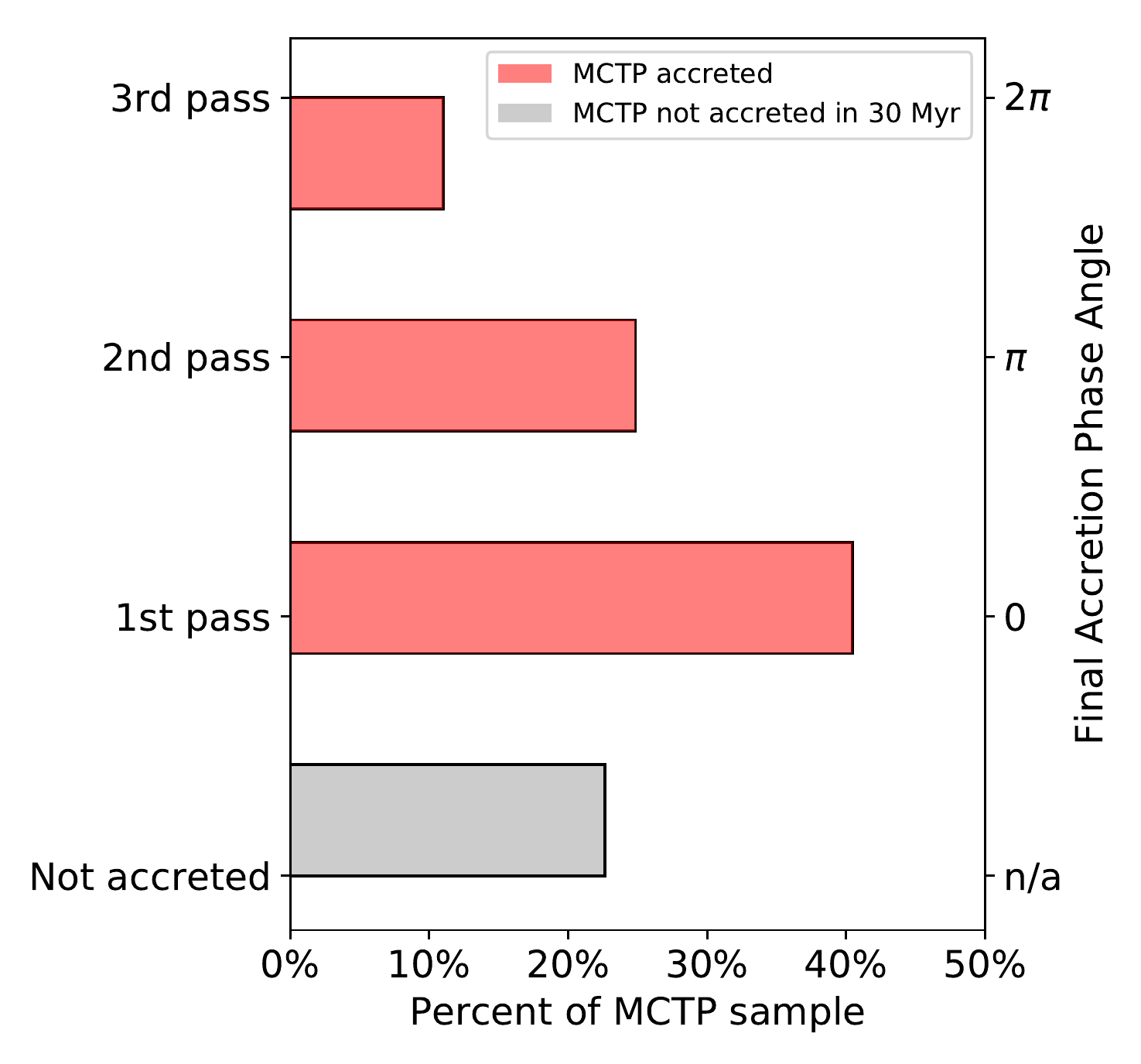}
\end{center}
\caption{The distribution of final accretion phase angles for the sample of 2$\times 10^4$ tracer particles selected from the near-side dust lane. The accretion phase angle is measured cumulatively from the the time at which the tracers are selected until the time at which the tracer enters a Galactocentric radius of 250 pc and stays within that radius for at least 4 Myr. About 40\% of the tracers are accreted upon their first pass (corresponding to a final accretion phase angle of $\sim 0$). Another $\sim$25\% of the total tracer sample is accreted upon the next pass, having transited $\pi$ rad, and $\sim$13\% more accrete again after the next pass, having transited 2$\pi$ rad. Of the initial sample, $\sim$23\% do not accrete within the amount of time simulated, either escaping out further into the disk or accreting at times past the limit of what was simulated. }
\label{fig:inflow_pangle}
\end{figure}

With the final accretion times of each tracer in hand, we can measure the number of ``passes'' by the the CMZ that each tracer takes to finally accrete. We measure this by the cumulative Galactocentric phase angle transited by each tracer particle. A histogram of the cumulative phase angle traveled before accretion is shown in Figure \ref{fig:inflow_pangle}. Of the initial sample of $2\times 10^4$ tracer particles, about 40\% accrete on the first pass (meaning their accretion phase angle is close to 0), then about 25\% of the remaining tracers accrete on the next pass. The number of tracers accreting on subsequent passes seems to diminish, as might be expected.

However, $\sim$23\% of the tracer particles do not accrete onto the CMZ within the span of time simulated. Of the tracers that are not yet accreted, virtually all remain in the bar region, and will be accreted at later times. While this makes it difficult to measure the ''average`` inflow efficiency using this method, it does allow us to place a lower limit on the number of passes, constraining the inflow efficiency. It appears from Figure \ref{fig:inflow_pangle} that between two and three overshooting events are typically required for the CMZ to accrete the majority of the tracer particles sampled. Because fewer than half of the tracer particles accrete on the first pass, the inflow efficiency cannot be greater than 0.5. 

If the tracer samples used in this experiment are characteristic of the dust lanes more generally, then we would expect an inflow efficiency of $\epsilon_{\text{DL}} \simeq 1/(\text{\# passes})$. Since more than two passes are required to accrete more than 75\% of the tracer particles (including those that do not accrete within the measured time), we know that the inflow efficiency is less than $\epsilon_{\text{DL}}\simeq 0.5$. While this is not a precise measurement of the efficiency, it does provides an intuitive upper bound and an interpretation of the more quantitatively rigorous estimates of $\epsilon_{\text{DL}}$ in Sections \ref{sec:inflow_bulk} and \ref{sec:inflow_inst}. 

\subsection{Complexities of Measuring the Inflow Efficiency}

The values for the inflow efficiency generated in the last three sections each have their own advantages and disadvantages. The bulk inflow efficiency derived in Section \ref{sec:inflow_bulk} is the simplest and most intuitive way of measuring what fraction of mass in the dust lanes actually accretes as a function of time, but averages away variations in the efficiency. The accretion phase angles derived in Section \ref{sec:inflow_history} permit us to understand more clearly how the overshooting gas re-accretes over time but is limited by the time window simulated, which would have to be very long to capture the accretion of the entire tracer sample. The instantaneous inflow efficiency derived in Section \ref{sec:inflow_inst} provides a middle ground, where we can measure the inflow efficiency for a large number of independent samples of tracer particles while observing their evolution for the requisite span of time to derive $\epsilon_{\text{DL}}$. In this section we address a couple of further complexities of the instantaneous inflow efficiency.

For instance, there is a possibility that $\epsilon_{\rm DL}$ may vary if we select smaller subsections of the dust lane. To address this, we determined the value of $\epsilon_{\rm DL}$ as a function of the length of the tracer selection region used. We found that the variations in $\epsilon_{\rm DL}$ resulting from changes to the size of the tracer selection region are small compared to the overall difference between the two dust lanes and the overall morphological variations of the dust lanes. In general, subdivided regions of tracers behave close to the mean behavior of the larger region that contains them. This may in part be due to the lack of self-gravity and supernova feedback which act to disrupt and fragment the infalling material. Feedback processes which result in clumpier streams of intermittent inflow might lead to a more variable $\epsilon_{\rm DL}$ for individual parcels of gas within the dust lanes, though the ``smooth'' dust lanes simulated in this work are able to reproduce a total inflow rate into the CMZ consistent with simulations that include star formation feedback and gas self-gravity (see \citealt{tress_Simulations_2020b}). 

Furthermore, the initial conditions of the gas in the disk and further out in the dust lanes may have a significant effect on the inflow rate because they determine the amount of gas available to flow along the dust lanes. Factors such as spiral arm structure, merger history, and any other changes to the gravitational potential may lead to different initial conditions for infalling gas, and therefore greater variations in the inflow rate. Quantifying these variations would require a dedicated experiment with variations to initial conditions of the simulation, which is an effort that is beyond the scope of this work.

As simulation time progresses, we observe changes in the effective impact parameter of the gas flowing along the dust lanes relative the x$_2$ orbits which have a time-varying effect on the inflow efficiency measured. The dust lanes appear to alternate between a state mostly intersecting the CMZ (with a higher-than-average inflow efficiency) and a state mostly overshooting (with a lower-than-average inflow efficiency). When the bar potential first matures ($t\approx 150$ Myr), most of the material in the dust lanes flows closer to the CMZ, whereas after about 5-10 Myr the impact parameter of most of the gas is larger and the majority of material overshoots. After another 10-15 Myr, the dust lanes again mostly intersect the CMZ, with a smaller impact parameter.

These observed variations are taken into account in the average value and remain in agreement with the reported uncertainty of $\epsilon_{\rm DL}$. We do not observe signatures of oscillations in the inflow efficiency on timescales greater than those resolved within the timescale of this simulation. While the instantaneous efficiency changes throughout the simulation due to the varying morphology of the dust lane region, the reported value for $\epsilon_{\rm DL}$ can be considered to be an average value for correcting the bulk inflow rate of the entire dust lane region.

\section{Following Gas on CMZ Orbits}
\label{sec:clouds}

\subsection{Cloud Identification and Tracing}\label{sec:cloud_ID}
In the previous sections, we characterized how gas gets accreted onto the CMZ within the simulated Galactic Center. We will now turn our attention to the behavior of clouds on orbits within the CMZ. Giant molecular clouds are hierarchical structures, and one's method of determining meaningful boundaries for such structures necessarily depends on the nature of the intended analysis. Within the simulated CMZ presented in this work, high density structures resembling molecular clouds form and evolve along their orbits of the Galactic Center. Overdensities that form in these clouds do not have the capacity to collapse past our imposed cell volume floor and they cannot form stars. With no feedback processes to disrupt the ISM, the resulting clouds have absolute masses and densities higher than would be expected of the astrophysical ISM. However, the lack of gas self-gravity and feedback allows us to directly investigate the isolated effects of the external potential on the dynamical evolution of the clouds. To this end, we must define a scheme to identify clouds and track their evolution in time.

\begin{figure}
\begin{center}
\includegraphics[trim = 20mm 0mm 0mm 0mm, width = .45\textwidth]{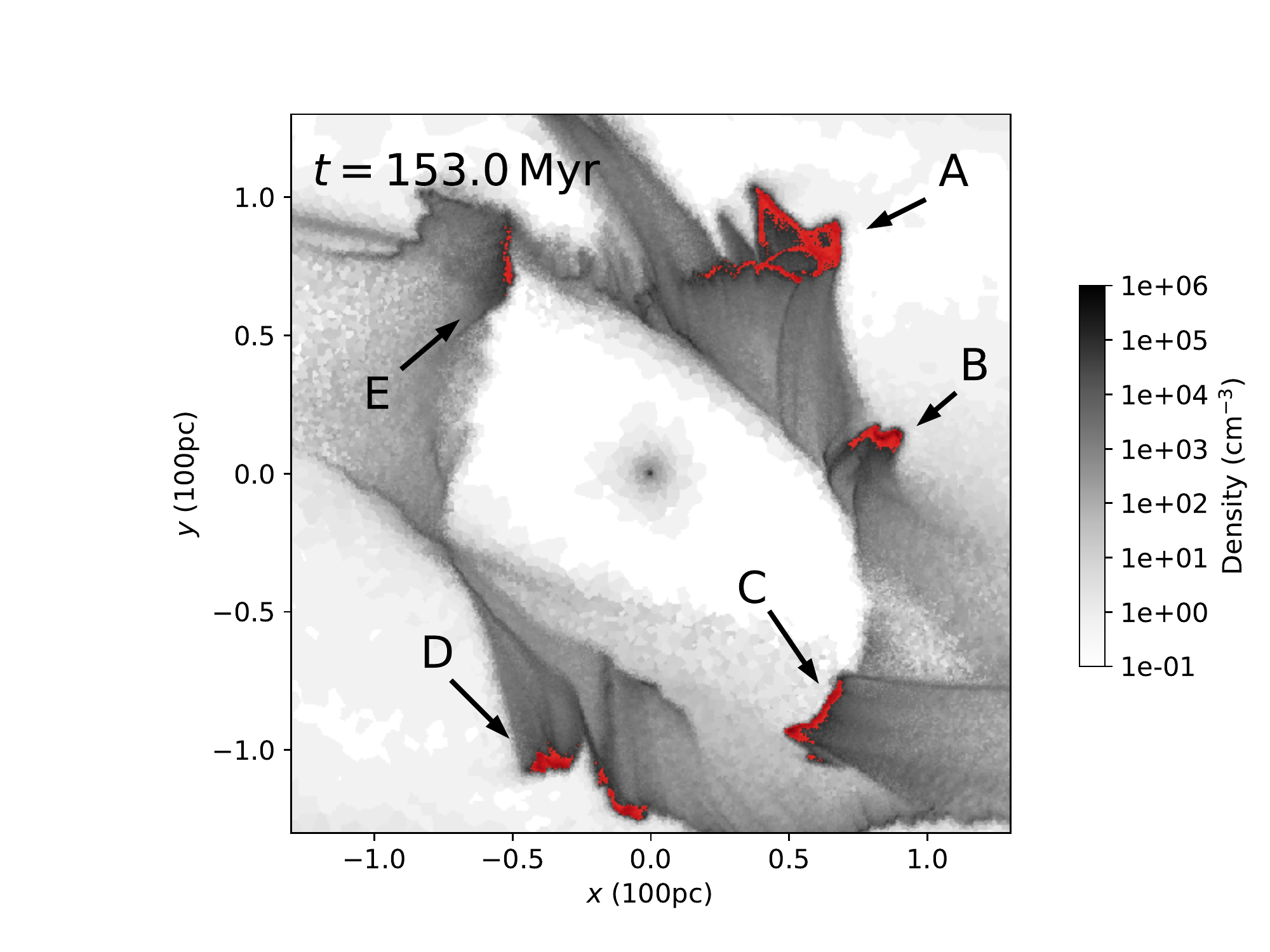}
\end{center}
\caption{A total gas density slice through $z=0$ of the simulated CMZ. The cloud structures discussed in Section \ref{sec:clouds} are highlighted in red as they are initially selected. The clouds are lettered for comparison with Figures \ref{fig:orbital_properties} and \ref{fig:phased_density}. The extent of the region illustrated here corresponds roughly to the dotted circle in Figure \ref{fig:cartoon}.}
\label{fig:cloud_intro}
\end{figure}

We adopt a simple criterion for cloud identification: a volume density boundary. Each corresponds to an isolated structure with a boundary satisfying a density requirement. In this context, ``isolated'' only means that the cloud is not subdivided into multiple separate components by the volume density boundary within the spatial extent initially chosen. We identify five such structures in the CMZ at a simulation time of 153 Myr using an initial volume density threshold of $10^{5}$~cm$^{-3}$. These clouds are lettered alphabetically (A-E), and are visualized in Figure \ref{fig:cloud_intro}. 

The spatial extent of each of these clouds is somewhat sensitive to these requirements. However, the mass-weighted mean cloud properties analyzed in this work are dominated by their high density components which are largely resilient to changes in this identification procedure (see Section \ref{sec:results:clouds}). 

At each timestep after tracer injection, we follow a simple procedure to determine which cells should be associated with the cloud in the next time step:
\begin{enumerate}
\itemsep0em
    \item Identify the unique IDs of tracer particles belonging to cells in the cloud in the present timestep. 
    \item In the next timestep, locate all of the cells who are parents of these tracers. A tracer's parent cell is the cell with which its position is associated at a given timestep.
    \item Check which of these parent cells satisfy the minimum volume density condition and a locality condition to prevent over-expansion of the cloud due to outliers.
    \item Label the cells that satisfy these conditions as associated with the cloud and record their properties.
    \item Repeat this process for the next timestep, and so on.
\end{enumerate}
This procedure allows the cloud to grow, shrink and change at each timestep, permitting new tracers to be advected into cloud-associated cells while other tracers may migrate far enough from the cloud to be excluded from the structure in future timesteps. The locality condition applied in the third step serves to exclude outlier tracers that happen to be advected rapidly away from the cloud association. These outliers are rare ($<0.1\%$ of the cloud's population) but if such a condition is neglected, or is too relaxed, then these outlier tracers cause the cloud to ``grow'' rapidly and in a way that is not representative of the bulk cloud motion. If the condition is too strict, the cells on the exterior of the cloud may be excluded at each timestep and the cloud may be forced to shrink at each timestep, which will lead to much smaller structures than the initially identified clouds. We impose a condition excluding all tracers hosted by cells with a radial distance (relative to the center-of-mass of the cloud) more than 50\% larger than the 2$\sigma$ radius that contains 95\% of the tracer particles. This allows the cloud to evolve, shrink and grow, while discarding statistically insignificant outlier tracers that would cause rapid over-expansion.

\subsection{Properties of Clouds on x$_2$ Orbits}\label{sec:results:clouds}
By considering the properties of our identified clouds as they evolve along their x$_2$ orbits in the absence of self-gravity, star formation and stellar feedback, we can glean some understanding of how CMZ clouds are affected by the external potential. While we emphasize again that these simulated structures may not consistently match the properties of observed Galactic Center clouds due to the limited physics included in the simulation, the observed dynamical effects ought to have a similar impact on real clouds subject to a similar external potential.

\begin{figure*}
\begin{center}
\includegraphics[trim = 0mm 0mm 0mm 0mm, width = .85 \textwidth]{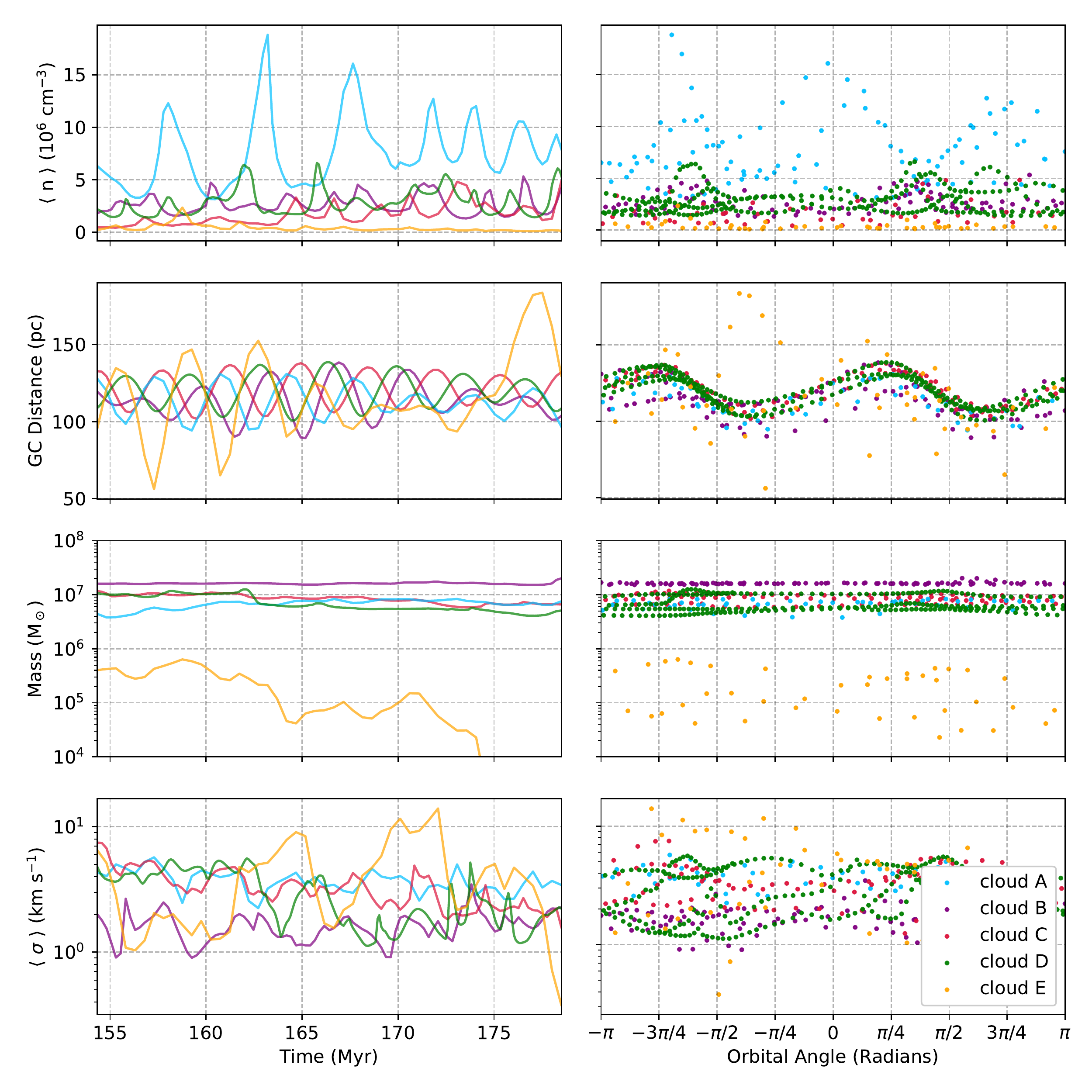}
\end{center}
\caption{The evolution of cloud properties for the five traced clouds highlighted in Figure \ref{fig:cloud_intro}. For each row, the left-hand panel shows the quantity's evolution as a function of time and the right-hand panel shows the same quantity as a function of orbital position-angle. The four rows display (from top to bottom) the cloud average number density, the Galactocentric radius of the cloud, the total cloud mass and finally the average velocity dispersion of the cloud. The apparent coincidence of density maxima with apocenter and pericenter passage (apocenter near an orbital phase of $-3\pi/4$ and $\pi/4$, and pericenter near $3\pi/4$ and $-\pi/4$) is quantified and analyzed in Section \ref{sec:discussion}}
\label{fig:orbital_properties}
\end{figure*}

In Figure \ref{fig:orbital_properties} we present the evolution of several key properties for the five simulated clouds as they progress along x$_2$ orbits. The clouds' densities seem to have sharply peaked maxima, varying significantly on a $\sim$1-2 Myr timescale. The Galactocentric radius of the clouds is plotted on the second row of panels to highlight the apocenter and pericenter passage of the clouds, which is discussed in more detail in Section \ref{sec:apocenter_passage}. 

The evolution of the total cloud masses is plotted in the third row, which shows that most of the clouds do not undergo rapid or substantial changes to their total mass as they orbit. Two of the clouds (D and E) seem to gradually lose mass, as their outermost material is stripped away by the powerful shear affecting clouds in the CMZ. This shear affects cloud E so severely that it eventually falls below the density criteria and effectively dissolves. The velocity dispersion of the clouds is calculated as the average standard deviation of the cloud velocity, $\langle \sigma\rangle = \text{std}(v_{\rm COM})$ where $v_{\rm COM}$ is the center of mass subtracted velocity of each cell and the angle brackets denote a mass-weighted average over each cell in the cloud. The evolution of this dispersion is plotted as a function of time and orbital phase in the bottom row of Figure \ref{fig:orbital_properties}. While the velocity dispersion of the clouds does vary, it does not seem to correlate with any of the other orbital properties. However, one should bear in mind that the dispersion would surely be strongly affected by the inclusion of gas self-gravity, star formation and feedback physics, all of which are absent in our simulations.

\subsection{Cloud Angular Momenta}\label{sec:methods:cloud_rotation}
The clouds identified in Section \ref{sec:cloud_ID} appear to have varying degrees of spin angular momentum, many showing signatures of counter-rotation with respect to the orbital angular momentum, while some appear to have short-lived co-rotation. To quantify this, we compare the significance of the cloud rotational motion to the velocity dispersion of the cloud using the following dimensionless ratio:
\begin{align}
\frac{\Omega_{\rm eff} R }{\sigma } =  \frac{1}{M_{\rm tot} R \sigma} \sum_i [m_i \vec{v}_i \times \vec{r}_{\rm i}]_z \,,
\end{align}
where $\Omega_{\rm eff}$ is an effective rotational velocity, $R$ is the cloud's effective radius, calculated as $R = \sqrt{\frac{1}{M_{\rm tot}} \sum_{i} m_i r_i^2}$ with $m_i$ and $r_i$ being the mass and center-of-mass radius of each cell composing the cloud respectively, $\sigma$ is the cloud's mean velocity dispersion, $\vec{v}_i$ is the velocity of the $i^{\rm th}$ particle in the cloud relative to the cloud center-of-mass rest frame, $\vec{r}_i$ is the $i^{\rm th}$ cell's position relative to the cloud center-of-mass, $z$ indicates the out-of-Galactic-plane axis, and $M_{\rm tot}$ is the summed mass of all cells composing the cloud. The velocity dispersion of each cloud, $\sigma$ is calculated as the Pythagorean sum of the standard deviations for the $x$, $y$, and $z$ components of the velocities in the center-of-mass rest frame, $\sigma = \sqrt{\sigma_x^2 + \sigma_y^2 + \sigma_z^2}$. A cloud that has $\text{abs}(\Omega_{\rm eff} R / \sigma) > 1$ displays ordered rotation which is more significant than the random motions within the cloud. Given the clockwise orbit for these clouds, a negative value for this ratio designates co-rotation and a positive value corresponds to counter-rotation. 

\begin{figure}
\begin{center}
\includegraphics[trim = 0mm 0mm 0mm 0mm, width = .4 \textwidth]{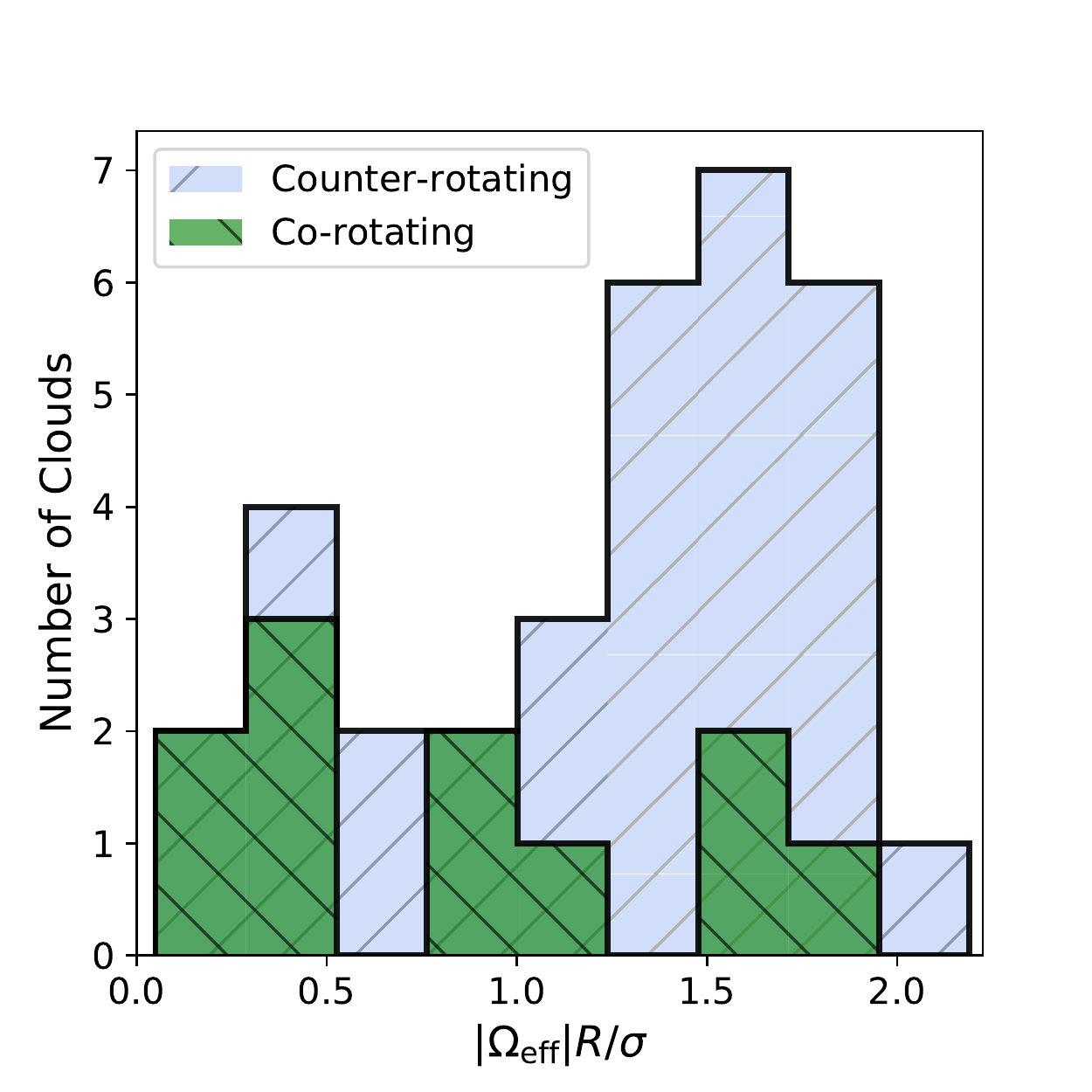}
\end{center}
\caption{A histogram of the significance of rotation, $\Omega_{\rm eff} R/\sigma$, for 44 clouds identified at different times throughout the simulation. The significance is calculated as the ratio of the effective rotational velocity of the cloud to its velocity dispersion, described in Section \ref{sec:rotation}. There are a greater number of counter-rotating clouds ($\sim 52\%$ of the sample), which is in agreement with previous studies of cloud rotation in the CMZ. However, significantly co-rotating clouds do occur, albeit much less frequently ($\sim 7\%$ of the sample).}
\label{fig:rotation_sample}
\end{figure}

We study the distribution of cloud angular momenta for a sample of 44 clouds isolated independently at different timesteps throughout the simulation. The results are presented in Figure \ref{fig:rotation_sample}. These clouds are selected using the same initial selection criteria used to identify the clouds in Section \ref{sec:cloud_ID}, but are not followed in time because the material at larger center-of-mass radii in each cloud contributes significantly to the overall angular momentum, and thus the resulting angular momenta are sensitive to the radial cutoff that is used to follow the clouds in time. Instead, we re-select the clouds at later times but do not follow the angular momentum evolution of each cloud individually. 

The majority of these clouds display significant counter-rotation, although a minority display co-rotation. Only 3 out of 44 clouds ($\sim 7\%$) show significant co-rotation, whereas 23 out of 44 ($\sim 52\%$) are significantly counter-rotating. The remainder have insignificant spin angular momentum. There is evidence for examples of counter-rotating clouds in the Milky Way's CMZ \citep[e.g.][]{federrath_link_2016}, but the spin angular momentum properties of such clouds are difficult to disentangle given their complex kinematic environment. Counter-rotation is a common feature observed in other recent simulations of the Galactic Center clouds \citep{dale_dynamical_2019,kruijssen_dynamical_2019}.

\subsection{Tracing the Origin of CMZ Clouds}\label{sec:methods:cloud_origin}
We also consider the origin of the mass composing a particular cloud, which is isolated using the same methods described in Section~\ref{sec:cloud_ID} but followed backwards in time. By flagging all of the tracers associated with this cloud and identifying the parent cells of those tracers at earlier time steps, we can track the origin and evolution of the distribution of mass which eventually composes the cloud. 

\begin{figure}
\begin{center}
\includegraphics[trim = 15mm 0mm 0mm 0mm, width = 0.48 \textwidth]{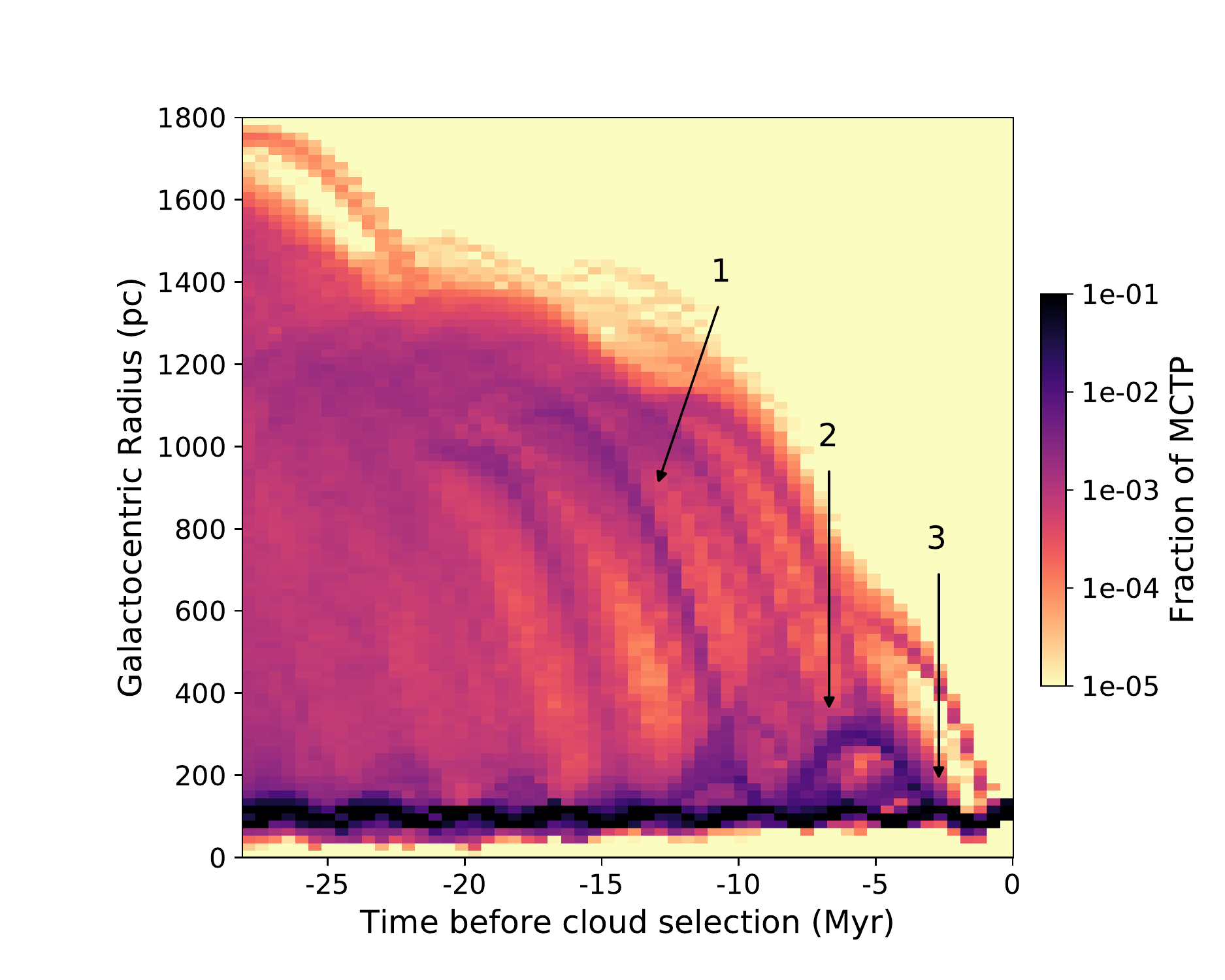}
\end{center}
\caption{The origin and evolution of mass tracers for a cloud identified late in the simulation, at $t \approx 178$~Myr. The intensity corresponds to the number of tracer particles that end up in the selected cloud as a function of time and Galactocentric radius. This distribution shows how a concentrated seed of the cloud persists within the CMZ (the high number of tracers oscillating around 100~pc in Galactocentric radius), while gradually accreting gas as it falls in along the dust lanes. A minimum of $\sim$10\% of the mass of the cloud originated outside the CMZ within the relevant timescale. Several features are numbered for comparison with their counterparts in other figures. Feature 1 indicates one of many instances of inflowing material that is destined to become part of the cloud's mass. In total, these inflows contribute $\sim 2 \times 10^6$M$_\odot$ to the cloud seed's mass over 10 Myr. Feature 2 highlights a portion of this inflowing material that overshoots, ``bouncing'' back to higher Galactocentric radii before merging with the cloud. Feature 3 indicates the re-accretion of the overshooting material as it combines with the cloud seed.}
\label{fig:origin_time}
\end{figure}

Figure \ref{fig:origin_time} shows the evolution of the tracers that will eventually constitute this cloud, binned as a function of time and Galactocentric radius (R$_{\rm GC}$). The cloud is selected from the snapshot corresponding to the right-most column of pixels at $t \approx 178$ Myr. At this time, the cloud is tightly localized at R$_{\rm GC} \approx 100$~pc, occupying a typical x$_2$ orbit. There is a concentration of tracer particles that primarily constitutes the cloud, remaining on its orbit for the duration of the simulation, while occasionally accreting the material streaming in from the dust lane. These accretion events are visible as the long arcs of tracer-density arriving from large values of R$_{\rm GC}$, labeled as feature 1. About a tenth of the cloud's future mass originates outside the CMZ just $10$ Myr before the cloud is selected, meaning a minimum $\sim$10\% of its tracers originate at R$_{\rm GC} > 250$~pc. Because the clouds produced in this simulation are more massive than their astrophysical counterparts, this fraction of mass amounts to $\sim 2\times 10^{6}$M$_{\odot}$ for this example cloud and should be thought of as a lower limit percentage for more realistic molecular clouds in the CMZ.

\section{Results and Discussion}
\label{sec:discussion}

\subsection{The Efficiency of Bar Driven Inflow}
A natural application for our estimate of the inflow efficiency is to calculate the mass accretion rate of the Milky Way's CMZ using observational estimates of the amount of mass within the Milky Way's dust lanes. Of the three methods that we have used to quantify the inflow efficiency, the instantaneous inflow efficiency calculated in Section \ref{sec:inflow_inst} best captures its variability as a function of time and dust lane morphology. Since all three methods agree within one standard deviation, we quote this instantaneous inflow efficiency for the purposes of correcting the observationally derived inflow rate from \cite{sormani_mass_2019}. Using this value of $\langle \epsilon_{\rm DL}\rangle$ we calculate the adjusted CMZ mass inflow rate as
\begin{equation}
    \dot{M}_{\rm CMZ} = \langle \epsilon_{\rm DL} \rangle \dot{M}_{\rm DL}
\end{equation}
where $\dot{M}_{\rm DL}$ is the total mass inflow rate under the assumption that the CMZ accretes all inflowing mass from the dust lanes, and $\langle \epsilon_{\rm DL}\rangle$ is the average dust lane accretion efficiency calculated in Section \ref{sec:inflow_inst}. 
We apply the observationally determined mass inflow rate of $\dot{M}_{\rm DL} = 2.7^{+1.5}_{-1.7}$ calculated by \citet{sormani_mass_2019}, which we use to find an adjusted CMZ mass inflow rate of $\dot{\text{M}}_{\rm CMZ}=0.8\pm 0.6$ M$_\odot$yr$^{-1}$. The errors in this value include uncertainties on the position-angle of the bar relative to the sun's position, the X$_{\rm CO}$ factor used to determine the mass content of the dust lanes, and the variability measured in this work for the dust lane inflow efficiency. The instantaneous value in the inflow efficiency for any given parcel of dust lane gas will be influenced by the initial conditions of inflowing gas, but the resulting average inflow rate is not strongly dependent on the inclusion of supernovae feedback or gas self-gravity \citep{tress_Simulations_2020b}. 

This CMZ mass inflow measurement solely accounts for the inflow rate from the dust lanes, and does not account for the loss of mass from supernova-driven outflows, which may drive winds out of the disk in the absence of AGN feedback \citep[e.g.\ ][]{crocker_NonThermal_2012,crocker_Unified_2015, armillotta_Life_2019}. We also do not account for the effects of the supermassive black hole Sgr A* \citep[e.g.\ ][]{su_giant_2010, wardle_origin_2014}, which may have, in the past, been a key mechanism for driving outflows from the Galactic Center.

Many models of the CMZ and its star formation rate depend upon the inflow rate \citep{longmore_Variations_2013, kruijssen_uncertainty_2014}, and in turn these models inform our understanding of how the Milky Way's stellar bulge evolves \citep{kormendy_Secular_2004}. The value calculated in this work is consistent with, but lower than previous measurements of the inflow rate, which was reported as $\dot{M}=1-3$~M$_\odot$yr$^{-1}$ by \citet{armillotta_Life_2019} and $\dot{M}\simeq 1 $~M$_\odot$yr$^{-1}$ by \citet{tress_Simulations_2020b}. Both of these previous measurements were performed using simulations including gas self-gravity, star formation, and feedback, suggesting that these processes do not affect the inflow efficiency considerably. The simulation used in this work to derive the inflow efficiency was tuned to replicate a Milky Way-like CMZ, but the prevalence of dust lanes observable in other barred spiral galaxies suggests that a similar efficiency might be applicable to other galaxies with similar morphology. A generalized understanding of how the inflow efficiency changes as a function of the parameters of the gravitational potential could be explored in future work, but lies beyond the scope of this paper.

\subsection{Accreted Gas is Well Distributed Across the CMZ}
\label{sec:inflow_distribution}
One of the consequences of the extreme shearing that occurs in the dust lanes and the partial overshooting of inflowing material is that a single parcel of gas in the dust lane will not be neatly deposited into an x$_2$ orbit as it accretes, nor will it be neatly integrated into a single existing CMZ cloud structure. Instead, the infalling material is well distributed across the CMZ, both due to shearing in the dust lane and the overshooting phenomenon. These effects can be seen in Figure~\ref{fig:inflow_snap_gallery}, where the tracer distribution becomes increasingly elongated and eventually fragments during its accretion. 

\begin{figure*}
\begin{center}
\includegraphics[trim = 0mm 0mm 0mm 0mm, width = 0.95 \textwidth]{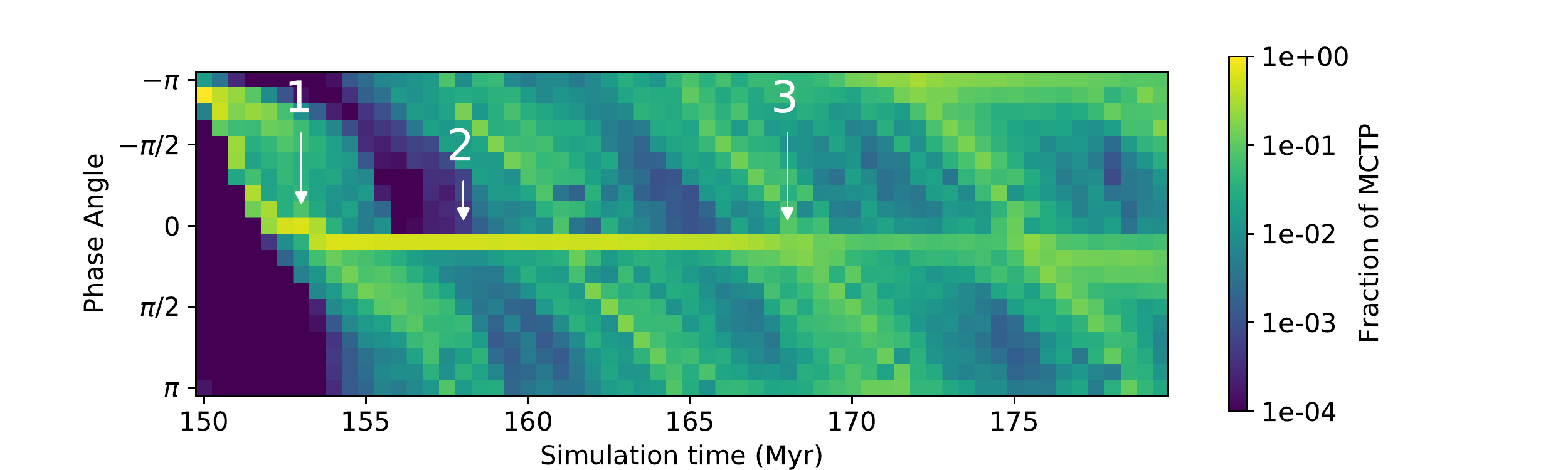}
\end{center}
\caption{The binned distribution of infalling material as a function of Galactocentric phase angle and time. The initial parcel of tracers is tightly packed in phase angle when it is selected at a simulation time of 150 Myr (labeled as feature 1), but is rapidly dispersed over the next 5~Myr. The portion of overshooting gas, labeled as feature 2, appears as the portion of tracers which remain at a constant position-angle of about 0 for the $\sim$15~Myr before re-accretion on the far-side of the CMZ. The re-accretion is labeled as feature 3.}
\label{fig:pangle_total}
\end{figure*}

We quantify the distributed accretion of gas by measuring the phase angle of tracers from a single parcel of gas in the dust lane. We select samples of tracers identically to criteria described in Section \ref{sec:inflow_inst}, and follow their motion. In Figure \ref{fig:pangle_total}, we show the time-evolving histogram of the tracer distribution as a function of Galactocentric phase angle as a proxy for how well the tracer particles have spread throughout the CMZ. The tracers initially occupy a narrow range of phase space, as the dust lane itself covers only a small position-angle. Upon approach, the distribution of tracers spreads out as the gas is sheared apart, entering the CMZ over an extended time. Star-forming clouds in an environment fed by such a mechanism may be continually fueled by this sheared out, accreting gas on timescales similar to the timescales on which they collapse and form stars ($\sim$ a few Myr). Additionally, the material that overshoots is largely re-accreted within the next 10-15~Myr. This infalling material is once again sheared apart upon its approach along the opposite dust lane. By the time it has been fully accreted, it is well distributed in position-angle. This re-accretion event is labeled with a (3) in Figures \ref{fig:inflow_percent}, \ref{fig:origin_time}, and \ref{fig:pangle_total}. 

This dynamically driven mechanism for mixing dust lane gas into the CMZ may influence the formation and evolution of molecular clouds, particularly the metallicity distribution of the Galactic Center. If individual structures in the dust lanes are sheared apart and accreted gradually over a large position-angle, then we would expect to see a CMZ that is continually fueled by material with a disk-like metallicity. This may influence the interpretation of the chemical composition and evolution of CMZ clouds \citep[e.g.\ ][]{tanaka_high_2011}.

\subsection{Apocenter Passage Drives Cloud Density Maxima}
\label{sec:apocenter_passage}

\begin{figure}
\begin{center}
\includegraphics[trim = 0mm 0mm 0mm 0mm, width = .48 \textwidth]{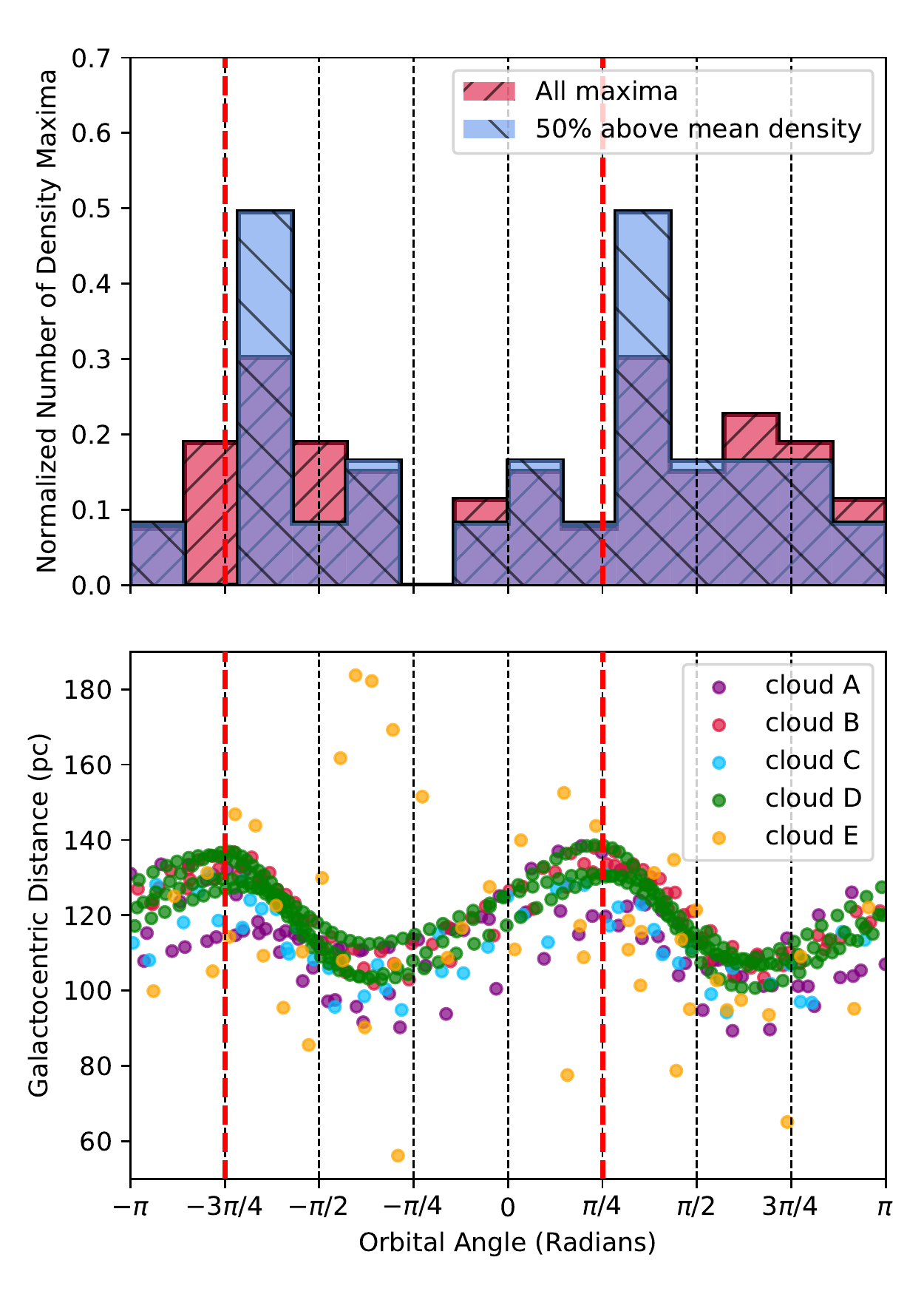}
\end{center}
\caption{Upper panel: the normalized histogram of local maxima in average cloud density as a function of orbital position-angle. Bottom panel: The phase angle evolution of Galactocentric radius for the five sample clouds identified in Section \ref{sec:clouds}. The distribution of the entire sample of density maxima (the red histogram) peak at an orbital position-angle corresponding to apocenter passage. This pattern is more pronounced if we only consider the local density maxima that exceed 50\% more than the mean cloud density (the blue histogram). The vertical dashed red lines indicate the position of typical apocenter passage in both panels. }
\label{fig:phased_density}
\end{figure}

Using the procedure discussed in Section \ref{sec:cloud_ID}, we are able to track the evolution of the long-lived clouds on CMZ x$_2$ orbits and investigate how their properties are affected by the dynamics of their environment. The density evolution of these clouds in Figure \ref{fig:orbital_properties} displays sharply peaked oscillations in the average cloud density on a timescale similar to the orbital period. The density increases do not consistently occur at the same position in orbital phase, but occur more frequently at the apocenter of the cloud's orbit. Figure \ref{fig:phased_density} shows the concentration of density maxima for each of the five clouds' evolution binned according to their orbital phase angle. A higher concentration of these significant peaks occur while the cloud is near apocenter, and a smaller number occur after pericenter. Many of these density fluctuations are relatively small. If we consider only the maxima that peak 50\% higher than the time-averaged cloud density (the blue histogram in Figure \ref{fig:phased_density}), then this pattern becomes more tightly focused on the orbit apocenter. The correlation of density maxima with orbital position is resilient to changes in the parameters of our cloud-following algorithm (e.g., as the cloud density criteria and the radial cutoff), as well as changes in the time resolution used between tracked snapshots. While the importance of pericenter passage and the density fluctuations that may be driven by that component of the orbit have been previously studied in the context of inducing star formation \citep{longmore_Candidate_2013,kruijssen_dynamical_2015, kruijssen_dynamical_2019, dale_dynamical_2019, jeffreson_physical_2018}, we find that apocenter passage plays a more significant role in pushing gas to higher densities and can potentially trigger gravitational collapse. 

Other properties of these clouds do not oscillate similarly. The total mass for each of the five clouds does not exhibit sharply peaked local maxima, instead they either remain relatively constant (e.g.\ cloud A in Figure \ref{fig:orbital_properties}) or they gradually change (e.g.\ clouds D and E in the same figure). The gradual decrease in mass observed in clouds D and E is due to the powerful shearing out of the clouds over time, while the few more sudden changes are due to cloud-cloud interactions and collisions with inflowing material. These effects are not sufficient to destroy the most of the cloud structures, with the exception of cloud E which dissipates due to shear over about 20 Myr. For more massive clouds, it appears that dynamics alone do not dominate cloud lifetimes in the CMZ. This result is in good agreement with recent analytical studies of cloud life cycles in this environment by \citet{jeffreson_general_2018} and \citet{jeffreson_physical_2018}, who find that gravitational collapse and star formation should be the dominant processes controlling cloud lifetimes in the CMZ. 

These density maxima are likely to be the result of multiple mechanisms. Firstly, as gas slows down near the apocenter, the clouds are compressed as in a traffic jam. However, if this were a uniform effect that occurs at each apocenter passage, then one should expect fluctuations in the density to occur consistently and have a similar magnitude at each apocenter passage. Because these average density peaks do not occur uniformly at each apocenter passage, this compression mechanism is insufficient to fully explain the observed behavior. A second mechanism driving these density fluctuations is that collisions with inflowing material tend to occur near the orbit apocenter. Gas accreting onto the CMZ displays inwards radial velocities ranging between 150 and 300 km s$^{-1}$. Generally, cloud-cloud collisions are capable of driving molecular clouds to high densities, and trigger gravitational collapse and efficient star formation, particularly in tumultuous environments such as the Galactic Center \citep[e.g][]{tsuboi_Cloud_2015,uehara_Molecular_2019}. The inflowing material is clumpy, and appears more clumpy in simulations that include self-gravity \citep{sormani_Simulations_2020, tress_Simulations_2020b}, which means that not every apocenter passage will guarantee a significant collision. This may help to explain the non-uniformity of the density fluctuations seen in Figure \ref{fig:orbital_properties}. 

\subsection{Counter-rotation and co-rotation of CMZ clouds}
\label{sec:rotation}
In Section \ref{sec:methods:cloud_rotation}, we find that a large portion of clouds in our simulation have significant angular momenta, most of them counter-rotating relative to the clockwise orbital angular momentum of the CMZ's x$_2$ orbits ($\sim 52\%$ counter-rotating, $\sim 7\%$ co-rotating, the rest with insignificant rotation). The rotation of molecular clouds \citep[e.g.][]{phillips_rotation_1999,imara_Angular_2011}, and more specifically the counter-rotation of clouds on CMZ orbits, is an expected result. This rotation is understood to be largely generated by the strong shear acting on clouds in this gravitational potential. However, this shear exclusively drives counter-rotation. Since the simulation does not include the effects of star formation and feedback that may play a role in disrupting cloud angular momenta in the astrophysical ISM, some other mechanism must be responsible for altering the angular momenta of these clouds to induce the infrequent co-rotation that we observe.

Each of the three most significant examples of co-rotating clouds (those with $|\Sigma_{\rm eff}| R/\sigma$>1) occur in the quarter orbit past apocenter passage, after a collision with inflowing material from the dust lanes. As inflowing gas from the dust lanes collides violently with gas at the apocenter of x$_2$ orbits, large quantities of angular momentum are transferred into CMZ clouds, generating short-lived co-rotation. The shear induced by the gravitational potential rapidly counteracts this short-lived co-rotation, and within a half-orbit the cloud displays either insignificant spin angular momentum or counter-rotation\footnote{Angular momentum is not a conserved quantity of a particle moving in a non-axisymmetric barred potential. Ultimately, all of the excess angular momentum of the gas is absorbed by the Galactic bar.}. Furthermore, the velocity gradients of these clouds might be significantly disrupted by physics not included in this work, such as the shocks from nearby supernovae or magnetic fields.

Some observed clouds in the CMZ show signs of counter-rotation, such as the rotation indicated by the velocity gradient of ``The Brick'' \citep{rathborne_G0_2014,federrath_link_2016,kruijssen_dynamical_2019}. The spin angular momenta for most CMZ clouds are not well constrained. It is possible that co-rotating clouds, resulting from inflow collisions, occur in the Milky Way's Galactic Center.

\subsection{The Bar Driven Feeding of CMZ Clouds}
While the shocks from supernovae and other stellar feedback effects are likely to influence the mixing of material within the CMZ, there is also a significant amount of mixing driven by dynamical forces. The strong shearing that is present for material on CMZ orbits and the interaction with the inflowing material on the dust lane shocks that are described in Section \ref{sec:methods:cloud_origin} lead to an environment where molecular clouds are constantly mixing with the surrounding gas, even in the absence of mixing effects from supernovae, due to the inflow from the dust lanes, which is in turn driven by the Galactic bar.

The time evolution of the tracers in Figure \ref{fig:origin_time} suggests a continually accreting cloud seed formed from the interaction of multiple CMZ clouds, as well as multiple distinct inflow events over a period of time of $\sim 3-4$ Myr, which is slightly longer than the expected collapse and star formation timescale suggested by \citealt{jeffreson_physical_2018}. If star formation in CMZ clouds are expected to collapse and form stars on timescales of $\sim$1-4 Myr, then the inflow driven by the bar will supply $\sim 0.8-5\times 10^5$M$_\odot$ of non-CMZ material to the clouds in the CMZ within that timescale. These accretion events occur near the apocenter of the cloud's orbit and may contribute fresh gas to sites of ongoing star formation, where winds from young stars may  already have blown away the natal gas cloud. 

\section{Summary}
\label{sec:summary}

We have performed hydrodynamic simulations of the Milky Way's CMZ with the aim of isolating the effects of the Milky Way's potential on gas inflow and properties of molecular clouds in this environment. We re-simulate the models of \cite{sormani_geometry_2019}, using a barred Milky Way's external potential with the addition of MCTP which advect between {\sc arepo}'s Voronoi cells proportional to the flow of mass, permitting us to quantitatively follow mass flows. Our main findings are:
\begin{itemize}
    \item We determine the inflow efficiency for gas accreting onto the CMZ from the bar dust lanes. This efficiency is found to be 30$\pm$12\%, and varies with fluctuations in the density and morphology of the dust lanes. We calculate the efficiency using three methods, all of which agree within the quoted variability. Using this, we calculate an adjusted value for the CMZ's observed inflow rate, originally calculated by \cite{sormani_mass_2019}, and find that $\dot{M}_{\rm cmz}=0.8\pm 0.6$ M$_{\odot}$ yr$^{-1}$.
    \item By following the evolution of clouds in the absence of self-gravity and star formation physics, we find that the barred potential drives sharp peaks in the average cloud density for CMZ clouds on x$_2$ orbits. These sudden changes to cloud density occur most frequently as the clouds pass their orbit's apocenter. This may be due to a combination of the orbital properties, as well as collisions with inflowing gas from the dust lanes, which intersect the x$_2$ orbits at their apocenter.
    \item In the absence of self-gravity and star formation physics, clouds in the Galactic Center have significant spin angular momentum oriented such that they counter-rotate. This effect has been noted previously and is attributed to the strong shear acting on CMZ clouds. However, we note the presence of short-lived co-rotation in a small subset of clouds (only three co-rotating clouds in a 44 cloud sample, or $\sim 7\%$). This co-rotation seem to be driven by high-velocity collisions with inflowing material because these rare co-rotating clouds occur within the quarter-orbit after apocenter passage.
    \item We find that, due to both the partial overshooting and the powerful shear within the dust lane, the inflowing material is rapidly distributed across the entire CMZ within $\sim$4 Myr. This same effect also informs the origin of the material constituting clouds in the Galactic Center, which are constantly exchanging mass with their surroundings and have a steady diet of fresh gas from the dust lanes. Using our determined inflow rate, we find that CMZ clouds are supplied $\sim 0.8-5.5 \times 10^6$M$_\odot$ of gas from the dust lanes within a 4 Myr timescale. This could help to inform the interpretations of metallicity distribution evolution of gas and stars in the CMZ, and suggests that stellar nurseries could efficiently gain fuel for further star formation during and well after the initial onset of gravitational collapse. 
\end{itemize}

These results are observed in a simulated ISM that isolates the effects of dynamical sources from the effects of self-gravity, star formation, and star formation feedback. The observed effects are largely driven by the external barred gravitational potential and by violent cloud-cloud collisions that occur as high-velocity gas from the dust lanes crashes into dense clouds in the CMZ near the apocenter of their orbits. These collisions may greatly impact the life cycle and evolution of molecular clouds in the CMZ.

\acknowledgments
We thank the anonymous referee for their helpful and constructive feedback that has improved this work. H.P.H. gratefully acknowledges support from the SOFIA Archival Research Program (program ID 09\_0540). H P.H. also thanks the LSSTC Data Science Fellowship Program, which is funded by LSSTC, NSF Cybertraining grant \#1829740, the Brinson Foundation, and the Moore Foundation; his participation in the program has benefited this work. C.B. and H.P.H. gratefully acknowledge support from the National Science Foundation under Award No. 1816715. M.C. S., R.G.T., S.C.O.G., and R.S.K. acknowledge support from the German Research Foundation (DFG) via the collaborative research center (SFB 881, Project-ID 138713538) the Milky Way System (subprojects A1, B1, B2, and B8), from the Heidelberg Cluster of Excellence “STRUCTURES” in the framework of Germany’s Excellence Strategy (grant EXC2181/1, Project-ID 390900948), and from the ERC via the ERC Synergy Grant “ECOGAL” (grant 855130). The project made use of computing resources provided by the state of Baden-Württemberg through bwHPC and the German Research Foundation (DFG) through grant INST 35/1134-1 FUGG. Data are stored at SDS@hd supported by the Ministry of Science, Research and the Arts Baden-Württemberg (MWK) and the German Research Foundation (DFG) through grant INST 35/1314-1 FUGG. R.J.S. gratefully acknowledges an STFC Ernest Rutherford fellowship (grant ST/N00485X/1) and H.P.C. from the Durham DiRAC supercomputing facility (grants ST/P002293/1, ST/R002371/1, ST/S002502/1, and ST/R000832/1). This research made use of \texttt{astropy}, a community-developed core Python package for astronomy \citep{astropy:2013,astropy:2018} and \texttt{NumPy} \citep{van2011numpy}.

\bibliographystyle{yahapj}
\bibliography{tracer_lib_final}

\begin{thebibliography}{}
\providecommand\natexlab[1]{#1}
\providecommand\JournalTitle[1]{#1}

\bibitem[{{Armillotta} {et~al.}(2019){Armillotta}, {Krumholz}, {Di Teodoro}, \&
  {McClure-Griffiths}}]{armillotta_Life_2019}
{Armillotta}, L., {Krumholz}, M.~R., {Di Teodoro}, E.~M., \&
  {McClure-Griffiths}, N.~M. 2019,
  \href{http://dx.doi.org/10.1093/mnras/stz2880}{\JournalTitle{\mnras}, 490,
  4401}

\bibitem[{{Astropy Collaboration} {et~al.}(2013){Astropy Collaboration},
  {Robitaille}, {Tollerud}, {Greenfield}, {Droettboom}, {Bray}, {Aldcroft},
  {Davis}, {Ginsburg}, {Price-Whelan}, {Kerzendorf}, {Conley}, {Crighton},
  {Barbary}, {Muna}, {Ferguson}, {Grollier}, {Parikh}, {Nair}, {Unther},
  {Deil}, {Woillez}, {Conseil}, {Kramer}, {Turner}, {Singer}, {Fox}, {Weaver},
  {Zabalza}, {Edwards}, {Azalee Bostroem}, {Burke}, {Casey}, {Crawford},
  {Dencheva}, {Ely}, {Jenness}, {Labrie}, {Lim}, {Pierfederici}, {Pontzen},
  {Ptak}, {Refsdal}, {Servillat}, \& {Streicher}}]{astropy:2013}
{Astropy Collaboration}, {Robitaille}, T.~P., {Tollerud}, E.~J., {et~al.} 2013,
  \href{http://dx.doi.org/10.1051/0004-6361/201322068}{\JournalTitle{\aap},
  558, A33}

\bibitem[{{Astropy Collaboration} {et~al.}(2018){Astropy Collaboration},
  {Sip{\H{o}}cz}, {G{\"u}nther}, {Lim}, {Crawford}, {Conseil}, {Shupe},
  {Craig}, {Dencheva}, {Ginsburg}, {VanderPlas}, {Bradley},
  {P{\'e}rez-Su{\'a}rez}, {de Val-Borro}, {Paper Contributors}, {Aldcroft},
  {Cruz}, {Robitaille}, {Tollerud}, {Coordination Committee}, {Ardelean},
  {Babej}, {Bach}, {Bachetti}, {Bakanov}, {Bamford}, {Barentsen}, {Barmby},
  {Baumbach}, {Berry}, {Biscani}, {Boquien}, {Bostroem}, {Bouma}, {Brammer},
  {Bray}, {Breytenbach}, {Buddelmeijer}, {Burke}, {Calderone}, {Cano
  Rodr{\'\i}guez}, {Cara}, {Cardoso}, {Cheedella}, {Copin}, {Corrales},
  {Crichton}, {D{\textquoteright}Avella}, {Deil}, {Depagne}, {Dietrich},
  {Donath}, {Droettboom}, {Earl}, {Erben}, {Fabbro}, {Ferreira}, {Finethy},
  {Fox}, {Garrison}, {Gibbons}, {Goldstein}, {Gommers}, {Greco}, {Greenfield},
  {Groener}, {Grollier}, {Hagen}, {Hirst}, {Homeier}, {Horton}, {Hosseinzadeh},
  {Hu}, {Hunkeler}, {Ivezi{\'c}}, {Jain}, {Jenness}, {Kanarek}, {Kendrew},
  {Kern}, {Kerzendorf}, {Khvalko}, {King}, {Kirkby}, {Kulkarni}, {Kumar},
  {Lee}, {Lenz}, {Littlefair}, {Ma}, {Macleod}, {Mastropietro}, {McCully},
  {Montagnac}, {Morris}, {Mueller}, {Mumford}, {Muna}, {Murphy}, {Nelson},
  {Nguyen}, {Ninan}, {N{\"o}the}, {Ogaz}, {Oh}, {Parejko}, {Parley}, {Pascual},
  {Patil}, {Patil}, {Plunkett}, {Prochaska}, {Rastogi}, {Reddy Janga},
  {Sabater}, {Sakurikar}, {Seifert}, {Sherbert}, {Sherwood-Taylor}, {Shih},
  {Sick}, {Silbiger}, {Singanamalla}, {Singer}, {Sladen}, {Sooley},
  {Sornarajah}, {Streicher}, {Teuben}, {Thomas}, {Tremblay}, {Turner},
  {Terr{\'o}n}, {van Kerkwijk}, {de la Vega}, {Watkins}, {Weaver}, {Whitmore},
  {Woillez}, {Zabalza}, \& {Contributors}}]{astropy:2018}
{Astropy Collaboration}, {Price-Whelan}, A.~M., {Sip{\H{o}}cz}, B.~M.,
  {G{\"u}nther}, H.~M., {et~al.} 2018,
  \href{http://dx.doi.org/10.3847/1538-3881/aabc4f}{\JournalTitle{\aj}, 156,
  123}

\bibitem[{Binney {et~al.}(1997)Binney, Gerhard, \&
  Spergel}]{binney_photometric_1997}
Binney, J., Gerhard, O., \& Spergel, D. 1997,
  \href{http://dx.doi.org/10.1093/mnras/288.2.365}{\JournalTitle{\mnras}, 288,
  365}

\bibitem[{Binney {et~al.}(1991)Binney, Gerhard, Stark, Bally, \&
  Uchida}]{binney_Understanding_1991}
Binney, J., Gerhard, O.~E., Stark, A.~A., Bally, J., \& Uchida, K.~I. 1991,
  \href{http://dx.doi.org/10.1093/mnras/252.2.210}{\JournalTitle{MNRAS}, 252,
  210}

\bibitem[{{Bland-Hawthorn} \& Gerhard(2016)}]{bland-hawthorn_galaxy_2016}
{Bland-Hawthorn}, J., \& Gerhard, O. 2016,
  \href{http://dx.doi.org/10.1146/annurev-astro-081915-023441}{\JournalTitle{ARA\&A},
  54, 529}

\bibitem[{Blitz \& Spergel(1991)}]{blitz_direct_1991}
Blitz, L., \& Spergel, D.~N. 1991,
  \href{http://dx.doi.org/10.1086/170535}{\JournalTitle{\apj}, 379, 631}

\bibitem[{Crocker(2012)}]{crocker_NonThermal_2012}
Crocker, R.~M. 2012,
  \href{http://dx.doi.org/10.1111/j.1365-2966.2012.21149.x}{\JournalTitle{\mnras},
  423, 3512}

\bibitem[{Crocker {et~al.}(2015)Crocker, Bicknell, Taylor, \&
  Carretti}]{crocker_Unified_2015}
Crocker, R.~M., Bicknell, G.~V., Taylor, A.~M., \& Carretti, E. 2015,
  \href{http://dx.doi.org/10.1088/0004-637X/808/2/107}{\JournalTitle{ApJ}, 808,
  107}

\bibitem[{Crutcher {et~al.}(1996)Crutcher, Roberts, Mehringer, \&
  Troland}]{crutcher_zeeman_1996}
Crutcher, R.~M., Roberts, D.~A., Mehringer, D.~M., \& Troland, T.~H. 1996,
  \href{http://dx.doi.org/10.1086/310031}{\JournalTitle{\apj}, 462, L79}

\bibitem[{Dahmen {et~al.}(1998)Dahmen, Huttemeister, Wilson, \&
  Mauersberger}]{dahmen_Molecular_1998}
Dahmen, G., Huttemeister, S., Wilson, T.~L., \& Mauersberger, R. 1998,
  \JournalTitle{A\&A}, 331, 959

\bibitem[{Dale {et~al.}(2019)Dale, Kruijssen, \&
  Longmore}]{dale_dynamical_2019}
Dale, J.~E., Kruijssen, J. M.~D., \& Longmore, S.~N. 2019,
  \href{http://dx.doi.org/10.1093/mnras/stz888}{\JournalTitle{MNRAS}, 486,
  3307}

\bibitem[{Federrath(2015)}]{federrath_inefficient_2015}
Federrath, C. 2015,
  \href{http://dx.doi.org/10.1093/mnras/stv941}{\JournalTitle{\mnras}, 450,
  4035}

\bibitem[{Federrath \& Klessen(2012)}]{federrath_Star_2012}
Federrath, C., \& Klessen, R.~S. 2012,
  \href{http://dx.doi.org/10.1088/0004-637X/761/2/156}{\JournalTitle{ApJ}, 761,
  156}

\bibitem[{Federrath {et~al.}(2014)Federrath, Schr{\"o}n, Banerjee, \&
  Klessen}]{federrath_Modeling_2014}
Federrath, C., Schr{\"o}n, M., Banerjee, R., \& Klessen, R.~S. 2014,
  \href{http://dx.doi.org/10.1088/0004-637X/790/2/128}{\JournalTitle{ApJ}, 790,
  128}

\bibitem[{Federrath {et~al.}(2016)Federrath, Rathborne, Longmore, Kruijssen,
  Bally, Contreras, Crocker, Garay, Jackson, Testi, \&
  Walsh}]{federrath_link_2016}
Federrath, C., Rathborne, J.~M., Longmore, S.~N., {et~al.} 2016,
  \href{http://dx.doi.org/10.3847/0004-637X/832/2/143}{\JournalTitle{ApJ}, 832,
  143}

\bibitem[{{Fux}(1999)}]{fux_3d_1999}
{Fux}, R. 1999, \JournalTitle{\aap}, 345, 787

\bibitem[{Gatto {et~al.}(2017)Gatto, Walch, Naab, Girichidis, W{\"u}nsch,
  Glover, Klessen, Clark, Peters, Derigs, Baczynski, \&
  Puls}]{gatto_SILCC_2017}
Gatto, A., Walch, S., Naab, T., {et~al.} 2017,
  \href{http://dx.doi.org/10.1093/mnras/stw3209}{\JournalTitle{\mnras}, 466,
  1903}

\bibitem[{Genel {et~al.}(2013)Genel, Vogelsberger, Nelson, Sijacki, Springel,
  \& Hernquist}]{genel_Following_2013}
Genel, S., Vogelsberger, M., Nelson, D., {et~al.} 2013,
  \href{http://dx.doi.org/10.1093/mnras/stt1383}{\JournalTitle{MNRAS}, 435,
  1426}

\bibitem[{Ginsburg {et~al.}(2016)Ginsburg, Henkel, Ao, Riquelme, Kauffmann,
  Pillai, Mills, {Requena-Torres}, Immer, Testi, Ott, Bally, Battersby,
  Darling, Aalto, Stanke, Kendrew, Kruijssen, Longmore, Dale, Guesten, \&
  Menten}]{ginsburg_dense_2016}
Ginsburg, A., Henkel, C., Ao, Y., {et~al.} 2016,
  \href{http://dx.doi.org/10.1051/0004-6361/201526100}{\JournalTitle{A\&A},
  586, A50}

\bibitem[{Girichidis {et~al.}(2016)Girichidis, Walch, Naab, Gatto, W{\"u}nsch,
  Glover, Klessen, Clark, Peters, Derigs, \& Baczynski}]{girichidis_SILCC_2016}
Girichidis, P., Walch, S., Naab, T., {et~al.} 2016,
  \href{http://dx.doi.org/10.1093/mnras/stv2742}{\JournalTitle{\mnras}, 456,
  3432}

\bibitem[{{Glover} \& {Clark}(2012)}]{Glover_approximations_2012}
{Glover}, S. C.~O., \& {Clark}, P.~C. 2012,
  \href{http://dx.doi.org/10.1111/j.1365-2966.2011.20260.x}{\JournalTitle{\mnras},
  421, 116}

\bibitem[{Glover \& Mac~Low(2007)}]{glover_Simulating_2007}
Glover, S. C.~O., \& Mac~Low, M.-M. 2007,
  \href{http://dx.doi.org/10.1086/512227}{\JournalTitle{ApJ}, 659, 1317}

\bibitem[{Henshaw {et~al.}(2016)Henshaw, Longmore, Kruijssen, Davies, Bally,
  Barnes, Battersby, Burton, Cunningham, Dale, Ginsburg, Immer, Jones, Kendrew,
  Mills, Molinari, Moore, Ott, Pillai, Rathborne, Schilke, Schmiedeke, Testi,
  Walker, Walsh, \& Zhang}]{henshaw_molecular_2016}
Henshaw, J.~D., Longmore, S.~N., Kruijssen, J. M.~D., {et~al.} 2016,
  \href{http://dx.doi.org/10.1093/mnras/stw121}{\JournalTitle{MNRAS}, 457,
  2675}

\bibitem[{Henshaw {et~al.}(2019)Henshaw, Ginsburg, Haworth, Longmore,
  Kruijssen, Mills, Sokolov, Walker, Barnes, Contreras, Bally, Battersby,
  Beuther, Butterfield, Dale, Henning, Jackson, Kauffmann, Pillai, Ragan,
  Riener, \& Zhang}]{henshaw_Brick_2019}
Henshaw, J.~D., Ginsburg, A., Haworth, T.~J., {et~al.} 2019,
  \href{http://dx.doi.org/10.1093/mnras/stz471}{\JournalTitle{MNRAS}, 485,
  2457}

\bibitem[{Heyer \& Dame(2015)}]{Heyer_molecular_2015}
Heyer, M., \& Dame, T. 2015,
  \href{http://dx.doi.org/10.1146/annurev-astro-082214-122324}{\JournalTitle{ARA\&A},
  53, 583}

\bibitem[{Imara \& Blitz(2011)}]{imara_Angular_2011}
Imara, N., \& Blitz, L. 2011,
  \href{http://dx.doi.org/10.1088/0004-637X/732/2/78}{\JournalTitle{ApJ}, 732,
  78}

\bibitem[{Jeffreson \& Kruijssen(2018)}]{jeffreson_general_2018}
Jeffreson, S. M.~R., \& Kruijssen, J. M.~D. 2018,
  \href{http://dx.doi.org/10.1093/mnras/sty594}{\JournalTitle{MNRAS}, 476,
  3688}

\bibitem[{Jeffreson {et~al.}(2018)Jeffreson, Kruijssen, Krumholz, \&
  Longmore}]{jeffreson_physical_2018}
Jeffreson, S. M.~R., Kruijssen, J. M.~D., Krumholz, M.~R., \& Longmore, S.~N.
  2018, \href{http://dx.doi.org/10.1093/mnras/sty1154}{\JournalTitle{MNRAS},
  478, 3380}

\bibitem[{Kalberla \& Dedes(2008)}]{Kalberla_global_2008}
Kalberla, P. M.~W., \& Dedes, L. 2008,
  \href{http://dx.doi.org/10.1051/0004-6361:20079240}{\JournalTitle{A\&A}, 487,
  951}

\bibitem[{Kauffmann {et~al.}(2017)Kauffmann, Pillai, Zhang, Menten, Goldsmith,
  Lu, \& Guzm{\'a}n}]{kauffmann_galactic_2017}
Kauffmann, J., Pillai, T., Zhang, Q., {et~al.} 2017,
  \href{http://dx.doi.org/10.1051/0004-6361/201628088}{\JournalTitle{A\&A},
  603, A89}

\bibitem[{Kim {et~al.}(2011)Kim, Saitoh, Jeon, Figer, Merritt, \&
  Wada}]{kim_NUCLEAR_2011}
Kim, S.~S., Saitoh, T.~R., Jeon, M., {et~al.} 2011,
  \href{http://dx.doi.org/10.1088/2041-8205/735/1/L11}{\JournalTitle{ApJ}, 735,
  L11}

\bibitem[{Kim {et~al.}(2012)Kim, Seo, Stone, Yoon, \&
  Teuben}]{kim_central_2012}
Kim, W.-T., Seo, W.-Y., Stone, J.~M., Yoon, D., \& Teuben, P.~J. 2012,
  \href{http://dx.doi.org/10.1088/0004-637X/747/1/60}{\JournalTitle{ApJ}, 747,
  60}

\bibitem[{Klessen \& Glover(2016)}]{klessen_Physical_2016}
Klessen, R.~S., \& Glover, S. C.~O. 2016,
  \href{http://dx.doi.org/10.1007/978-3-662-47890-5_2}{in Star {{Formation}} in
  {{Galaxy Evolution}}: {{Connecting Numerical Models}} to {{Reality}}:
  {{Saas}}-{{Fee Advanced Course}} 43. {{Swiss Society}} for {{Astrophysics}}
  and {{Astronomy}}, ed. Y.~Revaz, P.~Jablonka, R.~Teyssier, \& L.~Mayer,
  Saas-{{Fee Advanced Course}}} ({Berlin, Heidelberg}: {Springer}), 85

\bibitem[{Kormendy \& Kennicutt(2004)}]{kormendy_Secular_2004}
Kormendy, J., \& Kennicutt, R.~C. 2004,
  \href{http://dx.doi.org/10.1146/annurev.astro.42.053102.134024}{\JournalTitle{ARA\&A},
  42, 603}

\bibitem[{Krieger {et~al.}(2017)Krieger, Ott, Beuther, Walter, Kruijssen,
  Meier, Mills, Contreras, Edwards, Ginsburg, Henkel, Henshaw, Jackson,
  Kauffmann, Longmore, Martin, Morris, Pillai, Rickert, Rosolowsky, Shinnaga,
  Walsh, {Yusef-Zadeh}, \& Zhang}]{krieger_Survey_2017}
Krieger, N., Ott, J., Beuther, H., {et~al.} 2017,
  \href{http://dx.doi.org/10.3847/1538-4357/aa951c}{\JournalTitle{ApJ}, 850,
  77}

\bibitem[{Kruijssen {et~al.}(2015)Kruijssen, Dale, \&
  Longmore}]{kruijssen_dynamical_2015}
Kruijssen, J. M.~D., Dale, J.~E., \& Longmore, S.~N. 2015,
  \href{http://dx.doi.org/10.1093/mnras/stu2526}{\JournalTitle{MNRAS}, 447,
  1059}

\bibitem[{Kruijssen \& Longmore(2013)}]{kruijssen_Comparing_2013}
Kruijssen, J. M.~D., \& Longmore, S.~N. 2013,
  \href{http://dx.doi.org/10.1093/mnras/stt1634}{\JournalTitle{MNRAS}, 435,
  2598}

\bibitem[{Kruijssen \& Longmore(2014)}]{kruijssen_uncertainty_2014}
---. 2014, \href{http://dx.doi.org/10.1093/mnras/stu098}{\JournalTitle{MNRAS},
  439, 3239}

\bibitem[{Kruijssen {et~al.}(2019)Kruijssen, Dale, Longmore, Walker, Henshaw,
  Jeffreson, Petkova, Ginsburg, Barnes, Battersby, Immer, Jackson, Keto,
  Krieger, Mills, {S{\'a}nchez-Monge}, Schmiedeke, Suri, \&
  Zhang}]{kruijssen_dynamical_2019}
Kruijssen, J. M.~D., Dale, J.~E., Longmore, S.~N., {et~al.} 2019,
  \href{http://dx.doi.org/10.1093/mnras/stz381}{\JournalTitle{MNRAS}, 484,
  5734}

\bibitem[{Launhardt {et~al.}(2002)Launhardt, Zylka, \&
  Mezger}]{Launhardt_nuclear_2002}
Launhardt, R., Zylka, R., \& Mezger, P.~G. 2002,
  \href{http://dx.doi.org/10.1051/0004-6361:20020017}{\JournalTitle{A\&A}, 384,
  112}

\bibitem[{{Li} {et~al.}(2021){Li}, {Shen}, {Gerhard}, \& {Clarke}}]{Li+2021}
{Li}, Z., {Shen}, J., {Gerhard}, O., \& {Clarke}, J.~P. 2021,
  \JournalTitle{arXiv e-prints}, arXiv:2103.10342

\bibitem[{{Li} {et~al.}(2015){Li}, {Shen}, \& {Kim}}]{li_hydrodynamical_2015}
{Li}, Z., {Shen}, J., \& {Kim}, W.-T. 2015,
  \href{http://dx.doi.org/10.1088/0004-637X/806/2/150}{\JournalTitle{\apj},
  806, 150}

\bibitem[{{Liszt}(2008)}]{Liszt2008}
{Liszt}, H.~S. 2008,
  \href{http://dx.doi.org/10.1051/0004-6361:200809748}{\JournalTitle{\aap},
  486, 467}

\bibitem[{Longmore {et~al.}(2013{\natexlab{a}})Longmore, Kruijssen, Bally, Ott,
  Testi, Rathborne, Bastian, Bressert, Molinari, Battersby, \&
  Walsh}]{longmore_Candidate_2013}
Longmore, S.~N., Kruijssen, J. M.~D., Bally, J., {et~al.} 2013{\natexlab{a}},
  \href{http://dx.doi.org/10.1093/mnrasl/slt048}{\JournalTitle{\mnras}, 433,
  L15}

\bibitem[{Longmore {et~al.}(2013{\natexlab{b}})Longmore, Bally, Testi, Purcell,
  Walsh, Bressert, Pestalozzi, Molinari, Ott, Cortese, Battersby, Murray, Lee,
  Kruijssen, Schisano, \& Elia}]{longmore_Variations_2013}
Longmore, S.~N., Bally, J., Testi, L., {et~al.} 2013{\natexlab{b}},
  \href{http://dx.doi.org/10.1093/mnras/sts376}{\JournalTitle{MNRAS}, 429, 987}

\bibitem[{Marshall {et~al.}(2008)Marshall, Fux, Robin, \&
  Reyle}]{marshall_large_2008}
Marshall, D.~J., Fux, R., Robin, A.~C., \& Reyle, C. 2008,
  \href{http://dx.doi.org/10.1051/0004-6361:20078967}{\JournalTitle{A\&A}, 477,
  L21}

\bibitem[{Martini {et~al.}(2003{\natexlab{a}})Martini, Regan, Mulchaey, \&
  Pogge}]{Martini_circumnuclear_2003a}
Martini, P., Regan, M.~W., Mulchaey, J.~S., \& Pogge, R.~W. 2003{\natexlab{a}},
  \href{http://dx.doi.org/10.1086/367817}{\JournalTitle{\apjs}, 146, 353}

\bibitem[{Martini {et~al.}(2003{\natexlab{b}})Martini, Regan, Mulchaey, \&
  Pogge}]{martini_Circumnuclear_2003}
---. 2003{\natexlab{b}},
  \href{http://dx.doi.org/10.1086/374685}{\JournalTitle{ApJ}, 589, 774}

\bibitem[{McKee \& Ostriker(2007)}]{mckee_Theory_2007}
McKee, C.~F., \& Ostriker, E.~C. 2007,
  \href{http://dx.doi.org/10.1146/annurev.astro.45.051806.110602}{\JournalTitle{ARA\&A},
  45, 565}

\bibitem[{McMillan(2017)}]{mcmillan_mass_2017}
McMillan, P.~J. 2017,
  \href{http://dx.doi.org/10.1093/mnras/stw2759}{\JournalTitle{MNRAS}, 465, 76}

\bibitem[{Mills {et~al.}(2018)Mills, Ginsburg, Immer, Barnes, Wiesenfeld,
  Faure, Morris, \& {Requena-Torres}}]{mills_Dense_2018}
Mills, E. A.~C., Ginsburg, A., Immer, K., {et~al.} 2018,
  \href{http://dx.doi.org/10.3847/1538-4357/aae581}{\JournalTitle{ApJ}, 868, 7}

\bibitem[{Mills \& Morris(2013)}]{mills_detection_2013}
Mills, E. A.~C., \& Morris, M.~R. 2013,
  \href{http://dx.doi.org/10.1088/0004-637X/772/2/105}{\JournalTitle{ApJ}, 772,
  105}

\bibitem[{Morris \& Serabyn(1996)}]{morris_GALACTIC_1996}
Morris, M., \& Serabyn, E. 1996,
  \href{http://dx.doi.org/10.1146/annurev.astro.34.1.645}{\JournalTitle{ARA\&A},
  34, 645}

\bibitem[{Nelson \& Langer(1997)}]{Nelson_dynamics_1997}
Nelson, R.~P., \& Langer, W.~D. 1997,
  \href{http://dx.doi.org/10.1086/304167}{\JournalTitle{ApJ}, 482, 796}

\bibitem[{Orr {et~al.}(2021)Orr, Hatchfield, Battersby, Hayward, Hopkins,
  Wetzel, Benincasa, Loebman, Sormani, \& Klessen}]{orr_fiery_2021a}
Orr, M.~E., Hatchfield, H.~P., Battersby, C., {et~al.} 2021,
  \href{http://dx.doi.org/10.3847/2041-8213/abdebd}{\JournalTitle{\apj}, 908,
  L31}

\bibitem[{Phillips(1999)}]{phillips_rotation_1999}
Phillips, J.~P. 1999,
  \href{http://dx.doi.org/10.1051/aas:1999137}{\JournalTitle{A\&AS}, 134, 241}

\bibitem[{Pillai {et~al.}(2015)Pillai, Kauffmann, Tan, Goldsmith, Carey, \&
  Menten}]{pillai_Magnetic_2015}
Pillai, T., Kauffmann, J., Tan, J.~C., {et~al.} 2015,
  \href{http://dx.doi.org/10.1088/0004-637X/799/1/74}{\JournalTitle{ApJ}, 799,
  74}

\bibitem[{Portail {et~al.}(2017)Portail, Gerhard, Wegg, \&
  Ness}]{portail_dynamical_2017}
Portail, M., Gerhard, O., Wegg, C., \& Ness, M. 2017,
  \href{http://dx.doi.org/10.1093/mnras/stw2819}{\JournalTitle{\mnras}, 465,
  1621}

\bibitem[{Rahner {et~al.}(2019)Rahner, Pellegrini, Glover, \&
  Klessen}]{rahner_warpfield_2019}
Rahner, D., Pellegrini, E.~W., Glover, S. C.~O., \& Klessen, R.~S. 2019,
  \href{http://dx.doi.org/10.1093/mnras/sty3295}{\JournalTitle{\mnras}, 483,
  2547}

\bibitem[{Rathborne {et~al.}(2014)Rathborne, Longmore, Jackson, Foster,
  Contreras, Garay, Testi, Alves, Bally, Bastian, Kruijssen, \&
  Bressert}]{rathborne_G0_2014}
Rathborne, J.~M., Longmore, S.~N., Jackson, J.~M., {et~al.} 2014,
  \href{http://dx.doi.org/10.1088/0004-637X/786/2/140}{\JournalTitle{ApJ}, 786,
  140}

\bibitem[{Regan {et~al.}(1997)Regan, Vogel, \& Teuben}]{Regan_mass_1997}
Regan, M.~W., Vogel, S.~N., \& Teuben, P.~J. 1997,
  \href{http://dx.doi.org/10.1086/310717}{\JournalTitle{\apj}, 482, L143}

\bibitem[{Ridley {et~al.}(2017)Ridley, Sormani, Tre{\ss}, Magorrian, \&
  Klessen}]{ridley_nuclear_2017}
Ridley, M., Sormani, M.~C., Tre{\ss}, R.~G., Magorrian, J., \& Klessen, R.~S.
  2017, \href{http://dx.doi.org/10.1093/mnras/stx944}{\JournalTitle{MNRAS},
  469, 2251}

\bibitem[{Rosen \& Krumholz(2020)}]{rosen_role_2020}
Rosen, A.~L., \& Krumholz, M.~R. 2020,
  \href{http://dx.doi.org/10.3847/1538-3881/ab9abf}{\JournalTitle{AJ}, 160, 78}

\bibitem[{Sanders {et~al.}(2019)Sanders, Smith, \&
  Evans}]{sanders_pattern_2019}
Sanders, J.~L., Smith, L., \& Evans, N.~W. 2019,
  \href{http://dx.doi.org/10.1093/mnras/stz1827}{\JournalTitle{\mnras}, 488,
  4552}

\bibitem[{Smith {et~al.}(2020)Smith, Tre{\ss}, Sormani, Glover, Klessen, Clark,
  Izquierdo, Cabral, \& Zucker}]{smith_Cloud_2020}
Smith, R.~J., Tre{\ss}, R.~G., Sormani, M.~C., {et~al.} 2020,
  \href{http://dx.doi.org/10.1093/mnras/stz3328}{\JournalTitle{MNRAS}, 492,
  1594}

\bibitem[{Sormani \& Barnes(2019)}]{sormani_mass_2019}
Sormani, M.~C., \& Barnes, A.~T. 2019,
  \href{http://dx.doi.org/10.1093/mnras/stz046}{\JournalTitle{MNRAS}, 484,
  1213}

\bibitem[{Sormani {et~al.}(2015{\natexlab{a}})Sormani, Binney, \&
  Magorrian}]{sormani_Gas_2015a}
Sormani, M.~C., Binney, J., \& Magorrian, J. 2015{\natexlab{a}},
  \href{http://dx.doi.org/10.1093/mnras/stv441}{\JournalTitle{MNRAS}, 449,
  2421}

\bibitem[{Sormani {et~al.}(2015{\natexlab{b}})Sormani, Binney, \&
  Magorrian}]{sormani_Gas_2015b}
---. 2015{\natexlab{b}},
  \href{http://dx.doi.org/10.1093/mnras/stv2067}{\JournalTitle{MNRAS}, 454,
  1818}

\bibitem[{Sormani {et~al.}(2018{\natexlab{a}})Sormani, Sobacchi, Fragkoudi,
  Ridley, Tre{\ss}, Glover, \& Klessen}]{sormani_Dynamical_2018}
Sormani, M.~C., Sobacchi, E., Fragkoudi, F., {et~al.} 2018{\natexlab{a}},
  \href{http://dx.doi.org/10.1093/mnras/sty2246}{\JournalTitle{MNRAS}, 481, 2}

\bibitem[{Sormani {et~al.}(2020)Sormani, Tress, Glover, Klessen, Battersby,
  Clark, Hatchfield, \& Smith}]{sormani_Simulations_2020}
Sormani, M.~C., Tress, R.~G., Glover, S. C.~O., {et~al.} 2020,
  \href{http://dx.doi.org/10.1093/mnras/staa1999}{\JournalTitle{MNRAS}, 497,
  5024}

\bibitem[{Sormani {et~al.}(2018{\natexlab{b}})Sormani, Tress, Ridley, Glover,
  Klessen, Binney, Magorrian, \& Smith}]{sormani_theoretical_2018}
Sormani, M.~C., Tress, R.~G., Ridley, M., {et~al.} 2018{\natexlab{b}},
  \href{http://dx.doi.org/10.1093/mnras/stx3258}{\JournalTitle{MNRAS}, 475,
  2383}

\bibitem[{Sormani {et~al.}(2019)Sormani, Tre{\ss}, Glover, Klessen, Barnes,
  Battersby, Clark, Hatchfield, \& Smith}]{sormani_geometry_2019}
Sormani, M.~C., Tre{\ss}, R.~G., Glover, S. C.~O., {et~al.} 2019,
  \href{http://dx.doi.org/10.1093/mnras/stz2054}{\JournalTitle{MNRAS}, 488,
  4663}

\bibitem[{Springel(2010)}]{springel_pur_2010}
Springel, V. 2010,
  \href{http://dx.doi.org/10.1111/j.1365-2966.2009.15715.x}{\JournalTitle{MNRAS},
  401, 791}

\bibitem[{Su {et~al.}(2010)Su, Slatyer, \& Finkbeiner}]{su_giant_2010}
Su, M., Slatyer, T.~R., \& Finkbeiner, D.~P. 2010,
  \href{http://dx.doi.org/10.1088/0004-637X/724/2/1044}{\JournalTitle{ApJ},
  724, 1044}

\bibitem[{Tanaka {et~al.}(2011)Tanaka, Oka, Matsumura, Nagai, \&
  Kamegai}]{tanaka_high_2011}
Tanaka, K., Oka, T., Matsumura, S., Nagai, M., \& Kamegai, K. 2011,
  \href{http://dx.doi.org/10.1088/2041-8205/743/2/L39}{\JournalTitle{ApJ}, 743,
  L39}

\bibitem[{Tress {et~al.}(2020{\natexlab{a}})Tress, Smith, Sormani, Glover,
  Klessen, Mac~Low, \& Clark}]{tress_Simulations_2020a}
Tress, R.~G., Smith, R.~J., Sormani, M.~C., {et~al.} 2020{\natexlab{a}},
  \href{http://dx.doi.org/10.1093/mnras/stz3600}{\JournalTitle{MNRAS}, 492,
  2973}

\bibitem[{Tress {et~al.}(2020{\natexlab{b}})Tress, Sormani, Glover, Klessen,
  Battersby, Clark, Hatchfield, \& Smith}]{tress_Simulations_2020b}
Tress, R.~G., Sormani, M.~C., Glover, S. C.~O., {et~al.} 2020{\natexlab{b}},
  \href{http://dx.doi.org/10.1093/mnras/staa3120}{\JournalTitle{\mnras}, 499,
  4455}

\bibitem[{Tsuboi {et~al.}(2015)Tsuboi, Miyazaki, \& Uehara}]{tsuboi_Cloud_2015}
Tsuboi, M., Miyazaki, A., \& Uehara, K. 2015,
  \href{http://dx.doi.org/10.1093/pasj/psv076}{\JournalTitle{\pasj}, 67}

\bibitem[{Uehara {et~al.}(2019)Uehara, Tsuboi, Kitamura, \&
  Miyawaki}]{uehara_Molecular_2019}
Uehara, K., Tsuboi, M., Kitamura, Y., \& Miyawaki, R. 2019,
  \href{http://dx.doi.org/10.3847/1538-4357/aafee7}{\JournalTitle{ApJ}, 872,
  121}

\bibitem[{Van Der~Walt {et~al.}(2011)Van Der~Walt, Colbert, \&
  Varoquaux}]{van2011numpy}
Van Der~Walt, S., Colbert, S.~C., \& Varoquaux, G. 2011,
  \JournalTitle{Computing in Science \& Engineering}, 13, 22

\bibitem[{Walch {et~al.}(2015)Walch, Girichidis, Naab, Gatto, Glover,
  W{\"u}nsch, Klessen, Clark, Peters, \& Baczynski}]{walch_SILCC_2015}
Walch, S.~K., Girichidis, P., Naab, T., {et~al.} 2015,
  \href{http://dx.doi.org/10.1093/mnras/stv1975}{\JournalTitle{\mnras}, 454,
  246}

\bibitem[{Wardle \& {Yusef-Zadeh}(2014)}]{wardle_origin_2014}
Wardle, M., \& {Yusef-Zadeh}, F. 2014,
  \href{http://dx.doi.org/10.1088/2041-8205/787/1/L14}{\JournalTitle{ApJ}, 787,
  L14}

\bibitem[{Wegg \& Gerhard(2013)}]{wegg_mapping_2013}
Wegg, C., \& Gerhard, O. 2013,
  \href{http://dx.doi.org/10.1093/mnras/stt1376}{\JournalTitle{\mnras}, 435,
  1874}

\bibitem[{Weinberger {et~al.}(2020)Weinberger, Springel, \&
  Pakmor}]{weinberger_Arepo_2020}
Weinberger, R., Springel, V., \& Pakmor, R. 2020,
  \href{http://dx.doi.org/10.3847/1538-4365/ab908c}{\JournalTitle{ApJS}, 248,
  32}

\end{thebibliography}

\end{document}